\documentclass[useAMS,usenatbib]{mnras}
\usepackage{graphicx}
\usepackage[cp1251]{inputenc}
\usepackage[english]{babel}
\usepackage{multirow}
\usepackage[dvipsnames]{color}
\usepackage{etoolbox}
\makeatletter
\newcount\c@additionalboxlevel
\newcount\c@maxboxlevel
\patchcmd\@combinedblfloats{\box\@outputbox}{%
    \stepcounter{additionalboxlevel}%
    \box\@outputbox
}{}{\errmessage{\noexpand\@combinedblfloats could not be patched}}

\AtBeginShipout{%
    \ifnum\value{additionalboxlevel}>\value{maxboxlevel}%
    \typeout{Warning: maxboxlevel might be too small, increase to %
        \the\value{additionalboxlevel}%
    }%
    \fi
    \@whilenum\value{additionalboxlevel}<\value{maxboxlevel}\do{%
        \typeout{* Additional boxing of page `\thepage'}%
        \setbox\AtBeginShipoutBox=\hbox{\copy\AtBeginShipoutBox}%
        \stepcounter{additionalboxlevel}%
    }%
    \setcounter{additionalboxlevel}{0}%
}
\makeatother

\newcommand{\OIII}{[O~{\sc iii}]}

\newcommand{\NII}{[N~{\sc ii}]}
\newcommand{\SII}{[S~{\sc ii}]}

\newcommand{\HII}{H~{\sc ii}}
\newcommand{\HI}{H~{\sc i}}

\newcommand{\Ha}{H$\alpha$}

\newcommand{\NIIHa}{[N{\sc ii}]6548,6584/H$\alpha$}

\newcommand{\kms}{\,\mbox{km}\,\mbox{s}^{-1}}

\defcitealias{sit15}{Paper~I}

\begin{document}
\title
[ISM kinematics and YSOs in the Cyg OB1 supershell] {Star-forming regions at the periphery of
the supershell surrounding the Cyg OB1 association. II. ISM kinematics and YSOs in the star
 cluster vdB~130 region}

\author[Sitnik et al.]{
   T.G.~Sitnik$^{1}$\thanks{E-mail: tat.sitnik2011@yandex.ru},
   O.V.~Egorov$^{1, 2}$,
   T.A.~Lozinskaya$^{1}$,
   A.V.~Moiseev$^{2,1}$, \newauthor
   A.M.~Tatarnikov$^{1}$,
   O.V.~Vozyakova$^{1}$,
   D.S.~Wiebe$^{3}$\\
 $^{1}$Lomonosov Moscow State University, Sternberg Astronomical Institute,
        Universitetsky pr. 13, Moscow 119234, Russia \\
 $^{2}$ Special Astrophysical Observatory, Russian Academy of Sciences, Nizhnij Arkhyz 369167, Russia \\
 $^{3}$ Institute of Astronomy (INASAN), Russian Academy of Sciences, Pyatnitskaya str. 48, Moscow 119017, Russia}

\date{Accepted 2019 Month 00. Received 2019 Month 00; in original
form 2018 October 15}

\pagerange{\pageref{firstpage}--\pageref{lastpage}} \pubyear{2019}

\maketitle

\label{firstpage}

\begin{abstract}

We present an observational study of small-scale feedback processes operating
in the star-forming region located in the wall of the expanding supershell
around the Cyg OB1 association. The interstellar gas and dust content and
pre-stellar populations in the vicinity of the open star cluster vdB~130 are
analysed based on new optical and IR observations performed with the 6-m (3D
spectroscopic mapping in the [S~\textsc{ii}] doublet) and 2.5-m (optical and
NIR images) Russian telescopes along with the archival data of
\textit{Spitzer} and \textit{Herschel} space telescopes. Analysing
ionized gas kinematics and emission spectra, we discovered a
compact region with supersonic motions. These motions may be caused either by
stellar wind, or a bipolar outflow from a protostellar disc. Young
stellar objects were identified and classified in the area under study. Two
star-forming regions were identified. One of them is a region of ongoing star formation in
the head of the molecular cloud observed there and another one is a burst of star formation
in the cloud tail.

\end{abstract}


\section{Introduction}

      In recent years, with the growing availability of sub-mm and IR observations
of high angular resolution and sensitivity, investigations of star formation
prerequisites attract significant attention. In particular, new observational
facilities have extended modern studies toward numerous extragalactic
star-forming regions. The available observations indicate that in many
galaxies short local starburst episodes (with a duration of about 10 Myr)
that manifest themselves as complexes of ionized gas reside mainly within
dense walls of supershells around OB associations and in walls of giant \HI\
structures (see e.g. \citealt{Egorov2018} and references therein).

Turbulence, feedback from outflows, shell collisions, supernova explosions,
expanding \HII\ regions and other factors work together sculpting
a fine structure of molecular clouds with filaments, pillars, and blobs, peppered
with young stellar objects (YSOs). Given this complexity, along with
large-scale studies of giant supershells in other galaxies, a detailed look
at some nearby objects can be useful for analysing the star formation processes.

The aim of this work is to continue the study of star-forming regions in
walls of an expanding supershell formed by the wind and the UV radiation of
stars in the Cyg OB1 association \citep{loz88, loz90, sak92, loz97, loz98},
using our new optical and IR observations of the region along with archival
observational data from the \textit{Spitzer} and \textit{Herschel} space
telescopes.  The  Cyg OB1 association includes at
least 50 OB stars \citep{Humpreys}, and the size of the supershell
driven by this association is $3\times4$ degrees.

\begin{figure}
\centering
\includegraphics[width=0.99\linewidth]{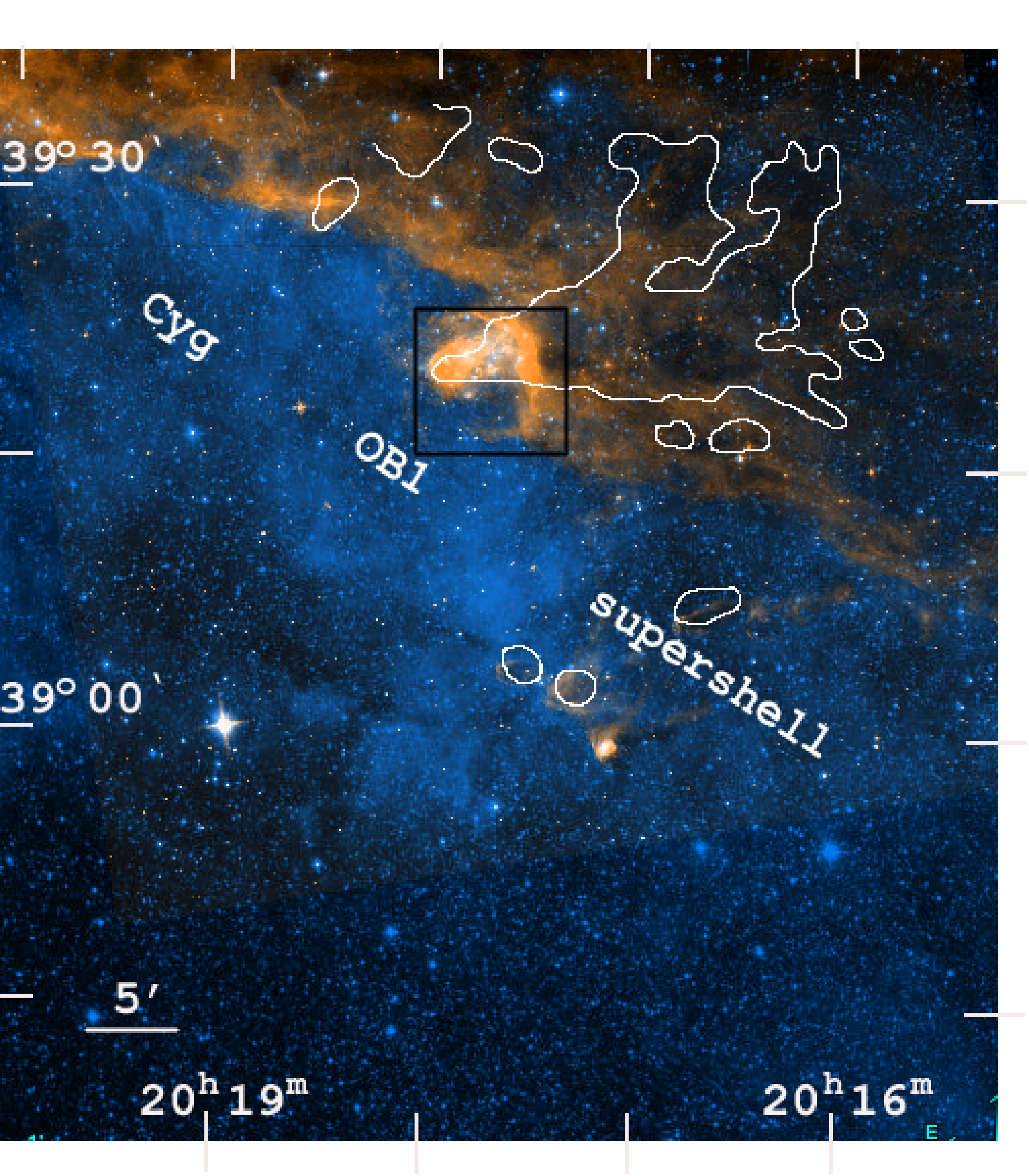}
\includegraphics[width=\linewidth]{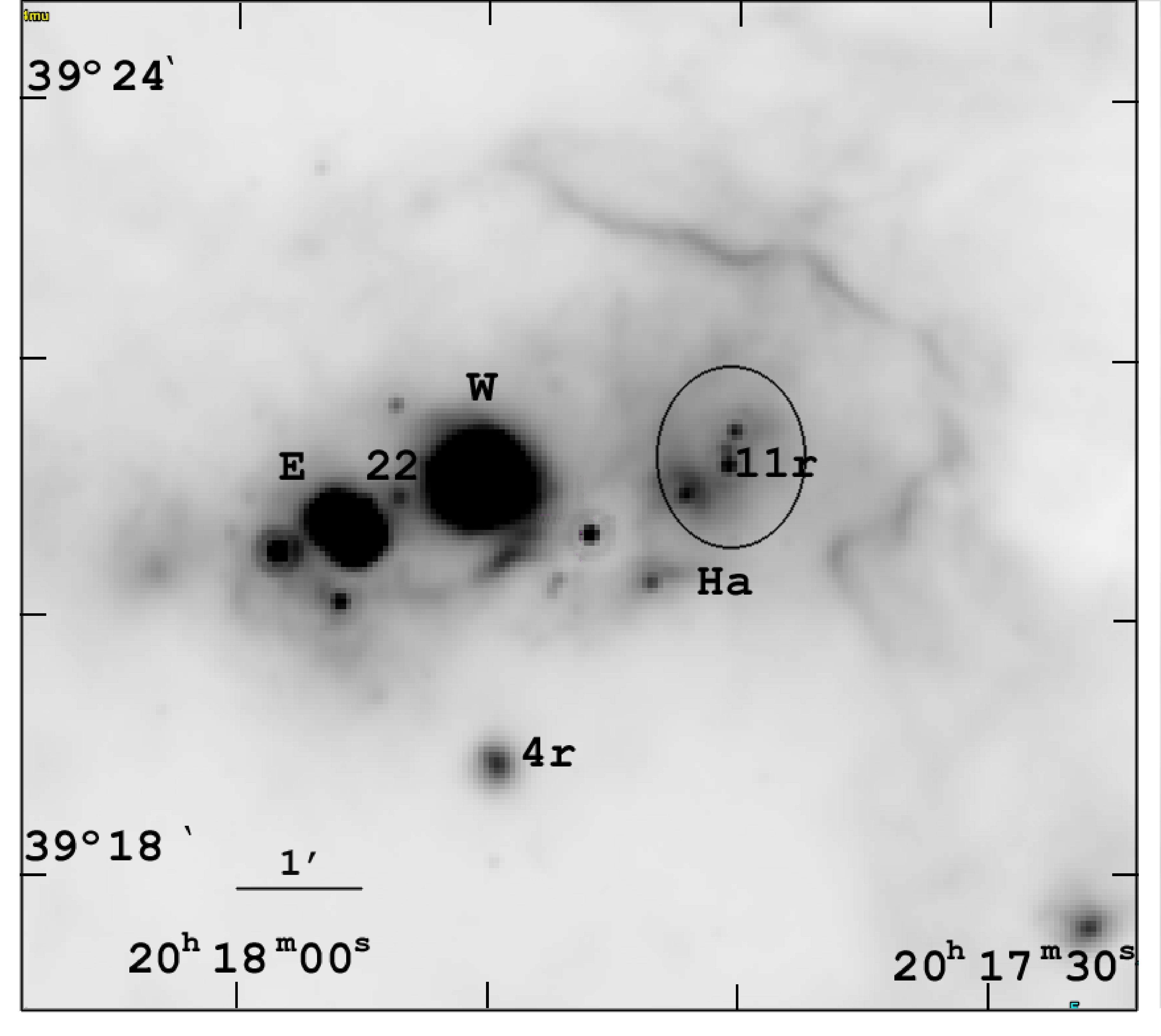}
\caption{The north-western part of the supershell around the Cyg OB1 stellar
association. { Shown is the combined image of the visual (blue) and IR (red) maps (based
on the {\em Spitzer} 5.8~$\mu$m band data) and CO  emission. A white contour shows the
approximate extent of the molecular cloud \citep{sch07}. Stars of Cyg OB1 are located
inside the supershell, in particular, in the lower left corner} (top panel). Shown
on the bottom panel is the zoomed {\em Spitzer} 24~$\mu$m band image of the
region marked by a square on the top panel. {Blob W, Blob E, \Ha\ Blob, and
several vdB~130 stars (4r, 11r and 22) discussed in \citetalias{sit15} are shown as well (see also Appendix)}.} \label{fig:vdB130}
\end{figure}

In \citet{sit15} (hereinafter referred to as Paper~I) and \citet{tatar16}
we have already partially studied the regions of ongoing star formation located in
the supershell and related to the globule family \citep{ark13} and the cometary
molecular cloud, or pillar \citep{sch07}. A young ($5-10$~Myr) star
cluster vdB 130 \citep{rac68} is observed in the direction of
the head of the CO cloud (see square on Fig.
\ref{fig:vdB130}, top). The physical relation between the Cyg OB1
association, the supershell, the vdB~130 cluster, and the cloud follows from
the distance and radial velocity estimates, as well as from several indirect
factors \citepalias[see][]{sit15}. With the assumed distance of $1.5-2$~kpc\footnote{{Further we use an
		average distance value of 1.7 kpc throughout the paper.}},
the size of the cluster, whose stars surround the head of the cometary cloud,
is no larger than 6~pc (10~arcmin) (Fig. \ref{fig:vdB130}, top). Some cluster stars
are surrounded by gas and dust nebulae that can be distinguished in the
optical and IR ranges. The central part of the cluster is enclosed in a
partial IR shell with a size of about 3~pc, visible in all \textit{Spitzer}
space telescope bands (Fig. \ref{fig:vdB130}, bottom). The interstellar
extinction law in the cluster region differs significantly from the normal
law, with $R_{v}=A_{v}/E(B-V)$ reaching values of $6-8$. On the other hand,
{the normal extinction law is valid along the line of sight to the cluster}
\citep{tatar16}. Some cluster stars are located in the centres of dust clumps
located in the interior of the IR shell and designated as Blob W and Blob E in \citetalias{sit15}.
Also, an \Ha\ emitting region referred to as `\Ha\ Blob' in \citetalias{sit15} resides
within the shell. The diameters of Blob W and Blob E are 0.4~pc and 0.2~pc,
respectively. Traces of a faint shock have been detected in the sky plane
between these blobs \citepalias{sit15}.

The structure of the supershell region under study is such that all the
extended sources (the cometary molecular cloud, the system of blobs) are
aligned in the direction of the illuminating source, that is, in the
direction of the closest projected Cyg~OB1 association stars, whereas the IR
shell around vdB~130 is opened toward the association.

In \citetalias{sit15} we considered the possibility of the following scenario
of the evolution of this region. In the cluster vicinity, the expanding
supershell around Cyg~OB1 interacts with the molecular cloud. A typical
cometary shape of the cloud (an IR pillar; see Fig.~\ref{fig:vdB130}, top) is
an indirect confirmation of such an interaction.According to this scenario,
the wind and the UV radiation of the Cyg OB1 stars had triggered the star
formation in the pillar, having resulted in the emergence of the vdB~130
cluster. Stars of the cluster interact with Blob W and Blob E in the east and
with the cometary cloud in the west, presumably, triggering the next episode
of the star formation. A similar scenario for the Carina nebula has been
discussed in \citet{smith}. To check the validity of this scenario for the IR
supershell, it would be good to confirm it with the age estimates as, for
example, in \cite{Oey2005}. Our estimate for the age of vdB 130 is in
apparent contradiction with the IR supershell age of 1 Myr given by
\citet{sak92}. However, we should keep in mind that the uncertainties in the
age determination are very large. So, the age comparison can be used neither
as an evidence in favour of triggering, nor as an argument against
triggering.

\begin{table*}
\caption{Summary of observational data} \label{tab:obs_data}
\begin{scriptsize}
\begin{tabular}{lccllcclcc}
\hline
Data set & Telescope,instrument & Filter  & Date of obs  & $T_{\rm exp}$, sec  & $\arcsec/$px  & $\theta$, $''$ &  sp. range  & $\lambda_C$, \AA & $\delta\lambda$ or FWHM, \AA \\
\hline
FPI Field 1 & 6-m, SCORPIO-2 & AC6730 & 17/18 Jun 2015 & $40\times60$ & 0.36  & 2.6 & \SII\, 6717, 6731\AA &6720 &$0.48 (22 \kms)$ \\
FPI Field 2 & 6-m, SCORPIO-2 &AC6730 & 17/18 Jun 2015 & $40\times60$ & 0.36  & 2.8 & \SII\, 6717, 6731\AA &6720 &$0.48 (22 \kms)$ \\
BTA images, field \#1 & 6-m, SCORPIO-2 &FN655 & 16/17 Oct 2014 & $2\times180$  & 0.36  &  3.5 &    \Ha$+$\NII+cont. &6559 &97 \\
                   & 6-m, SCORPIO-2 & FN674 & 16/17 Oct 2014 & $2\times180$ & 0.36 & 3.5 &     \SII+cont. &6733 &60 \\
                   & 6-m, SCORPIO-2 & FN641 & 16/17 Oct 2014 & $2\times80$   & 0.36  &  3.5 &     continuum &6413 &179 \\
                   & 6-m, SCORPIO-2 & FN712 & 16/17 Oct 2014 & $2\times80$   & 0.36  & 3.5 &     continuum &7137 &209 \\
BTA images, field \#2 & 6-m, SCORPIO-2 & FN655 & 16/17 Oct 2014 & $1\times180$  & 0.36  &  4.1 &    \Ha$+$\NII+cont. &6559 &97 \\
                   & 6-m, SCORPIO-2 & FN674 & 16/17 Oct 2014 & $1\times180$ & 0.36 & 4.1 &     \SII+cont. & 6733 &60 \\
                   & 6-m, SCORPIO-2 & FN641 & 16/17 Oct 2014 & $1\times80$  & 0.36  &  4.1 &     continuum & 6413 &179 \\
                   & 6-m, SCORPIO-2 & FN712 & 16/17 Oct 2014 & $1\times80$  & 0.36 & 4.1 &     continuum &7137 &209 \\
CMO image, field \#1 & 2.5-m, NBI & OIII & 26/27 Aug 2017& $2\times600$ &    0.16 & 2.3 & \OIII+cont. & 4992 & 63 \\
CMO image, field \#2 & 2.5-m, NBI & OIII & 26/27 Aug 2017 &  $2\times600$ &    0.16 & 2.7 & \OIII+cont. &4992 &63 \\
CMO image, field \#3 & 2.5-m, NBI & OIII & 27/28 Aug 2017 & $10\times300$ &    0.16 & 1.3 & \OIII+cont. &4992 &63 \\
CMO image, field \#4 & 2.5-m, NBI & OIII & 27/28 Aug 2017 & $7\times300$ &    0.16 & 1.4 & \OIII+cont. & 4992 &63 \\
CMO NIR Images & 2.5-m, ASTRONIRCAM & CO & 17/17 Dec 2017 & $30\times40$ &   0.27 & 0.8 & CO+cont.  & 2.285 $\mu$m  &302 \\
 & 2.5-m, ASTRONIRCAM & J & 07/08 Mar 2016 & $30\times100$ &  0.27 & 1.2 & continuum & 1.25 $\mu$m  &1665 \\
 & 2.5-m, ASTRONIRCAM & H & 07/08 Mar 2016 & $3\times20$ &   0.27 & 1.2 & continuum   &1.64 $\mu$m  &2928 \\
 & 2.5-m, ASTRONIRCAM & K & 26/27 Apr 2017 & $30\times40$ &   0.27 & 0.8 & continuum  & 2.2 $\mu$m  &3162 \\
  \hline
\end{tabular}
\end{scriptsize}
\end{table*}

The possibility of triggered star formation is a hot topic, which is widely
discussed in the literature, with numerous examples having been suggested
both for Galactic and extragalactic sources. However, the subject is very
uncertain, and various proposed signposts of the triggered star formation can
be insufficient. On the one hand, observations do indicate higher density of
younger stellar objects at edges of bubble structures around massive stars
\citep[see e. g.][]{2017A&A...605A..35P}. On the other hand, features that
are believed to be signatures of triggering can be misleading, even when they
are quite persuasive \citep{Dale}.

To study further this region, in this work we present new observational data,
which are used to analyse the ionized gas kinematics and the optical
 and IR emission-line morphology of the vdB~130 cluster region. These observations and methods of their reduction are described in Section~2. Section~3 describes the results of narrow-band optical and infrared observations. In Section~4 we analyse the kinematics of the area, and uncover traces of interaction between the stellar and interstellar populations. Possible causes of the appearance of supersonic motions are discussed in Section~5. An analysis of the distribution of young stellar objects in the area based on the observations of the
\textit{Spitzer} space telescope is presented in Section~6. The obtained results are summarized in Section~7.

\section{Observations and data reduction}

This study is based on optical observations performed with the 6-m telescope
of the Special Astrophysical Observatory of the Russian Academy of Sciences
(SAO RAS) and the 2.5-m telescope of the Caucasian Mountain Observatory (CMO)
of Sternberg Astronomical Institute, Lomonosov Moscow State University (SAI
MSU), as well as on archival IR data obtained with \textit{Spitzer} and
\textit{Herschel} space observatories. The log of our observations used in
this analysis, including those previously described in \citetalias{sit15}, is
given in Table~\ref{tab:obs_data}, where the used instrument and filter, the
total exposure time ($T_{\rm exp}$), the pixel size of the final images
($\arcsec/$px), the final angular resolution ($\theta$), the covered spectral
range, the central wavelength of the used filter ($\lambda_C$) and the final
spectral resolution ($\delta\lambda$ or FWHM) or the bandwidth of the used
filter (FWHM) are indicated for each data set.

\begin{figure*}
\includegraphics[width=0.9\linewidth]{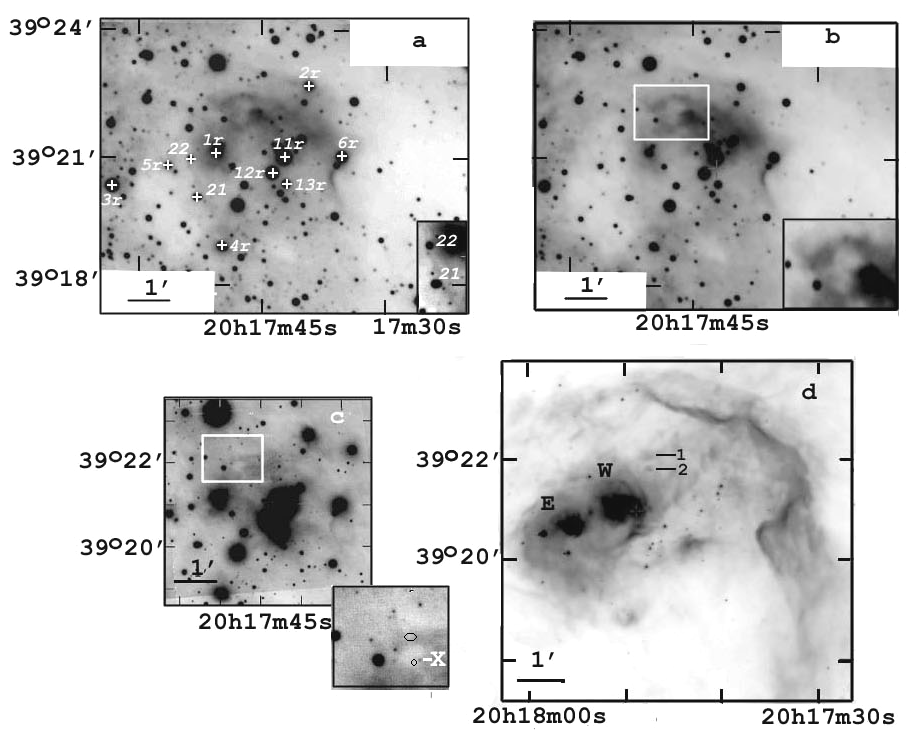}
\caption{{Morphology of the vicinity of the cluster vdB~130 as seen in \SII\ (a), \Ha\ (b), \OIII\ (c), and 8~$\mu$m bands presented on the same spatial scale.} 
Several stars
belonging to the cluster are denoted by numbers in panel {\it a}. The
fragment of the \SII-filament where the shock front propagates is shown in
the lower right corner. In lower right corners of panels {\it b} and {\it c}
zoomed images of the cavity region (indicated with rectangles) are shown
{(see discussion in Section 5)}. In panel {\it d} blobs E and W are
indicated and two IR features are marked by numbers and are also shown
{in the zoomed image in panel {\it c} with black ellipses}. The letter
X in panel {\it c} marks the point source in the direction of the cavity.}
\label{fig:Ha_Hb_SII_OIII}
\end{figure*}

\subsection{Fabry--Perot observations}

In order to investigate the ionized gas kinematics in the studied region we
observed it with a scanning Fabry--Perot interferometer (FPI)
mounted inside the SCORPIO-2 multi-mode focal reducer \citep{scorpio2} at
the 6-m telescope of SAO RAS. The operating spectral range around the \SII\,
emission lines was cut by a bandpass filter with an
$\mathrm{FWHM}\approx15$~\AA\ bandwidth. During the scanning process, we have
consecutively obtained 40 interferograms at different distances between the
FPI plates.

Data reduction was performed using a software package running in the
\textsc{idl} environment. For a detailed description of the data reduction
algorithms and software see \citet{Moiseev2002,  Moiseev2008} and
\citet{Moiseev2015}. After the initial reduction, sky line subtraction,
photometric and seeing corrections made using reference stars, and wavelength
calibration, the observational data were combined into data cubes, where each
pixel in the field of view contains a 40-channel spectrum. Because of the
large angular size of the region, we separated it into two overlapping
fields. Each of them was observed at two position angles in order to remove
the parasitic ghost reflection. These data were reduced separately to obtain
the wavelength cubes for each field. After heliocentric correction, we merged
both fields into single mosaic data cubes. In order to increase
signal-to-noise ratio we made $3\times3$ pixels binning of this mosaic data
cube to the pixel size of 1.05 arcsec. This spatial scale provides an optimal
sampling of data taken with a typical seeing value of 2.8 arcsec (see
Tab.~1).

The \SII\, line profiles in the obtained data cube were analysed using
single-component Voigt fitting \citep{Moiseev2008}. Because both \SII\,
6717~\AA\ and 6731~\AA\ (from neighbouring orders of interference) lines are
visible in the resulting spectra, we fitted them simultaneously with two
Voigt functions having similar FWHM and fixed velocity separation. To describe ionized gas motions in
several regions with line-profile asymmetry we used multi-component Voigt
fitting.

Unfortunately, weather conditions during the FPI observations were far
from ideal: the channel-to-channel seeing variations were about 20 per cent,
and the transparency variations were 20--40 per cent. We corrected this effect
using photometry of field stars as described in \citet{Moiseev2002}. However,
even after photometric corrections we detected artefacts in some regions
of the final data cube caused by the atmospheric transparency variations:
arc-like structures on the maps of emission lines parameters. The maximum
RMS of the parasitic arcs is 1.5 km/s in the velocity field and 6 km/s in the
velocity dispersion maps. These values limit the accuracy of our
line-of-sight velocity estimates presented below.

\subsection{Optical and NIR narrow-band images}

We obtained a large set of images in the vicinity of vdB~130 cluster in the following spectral bands: \Ha, \SII\, and \OIII\, optical emission lines; the optical continuum on both sides of \Ha\, and \SII\, lines; the NIR CO($v = 2-1$) emission line together with the J, H, and K band continuum. The parameters of all the data sets are given in Table~\ref{tab:obs_data}.

\Ha, \SII\, and optical continuum observations were performed with the
6-m BTA telescope of SAO RAS equipped with the SCORPIO-2 multi-mode focal
reducer. Because of the limited field of view ($6\times6$~arcmin), we
observed two overlapping fields separately, and then combined the reduced
images for each filter into a mosaic. The details of the observations and
reduction of these data sets are given in \citetalias{sit15}. Note that
because the FWHM of the used FN655 filter is broader than the separation
between \Ha\, and \NII\, emission lines, the image in this filter is
contaminated by \NII\, 6548, 6584~\AA\, emission. Given that the typical
ratio of \NIIHa\ $\simeq 0.4$ in the region {\citepalias{sit15}}, we
may expect the contribution of \NII\ lines to be up to 30~per cent of the
total observed flux. Hereafter, the `\Ha' notation is used to refer to the emission in the
\Ha\, line with this additional contribution due to the \NII\, lines.

The \OIII\, line images of vdB~130 vicinity were obtained with a
4k$\times$4k camera (made in the Niels Bohr Institute, Copenhagen; here and
after -- NBI camera) mounted in the Cassegrain focus of the 2.5-m telescope
CMO of SAI MSU. In this camera, the CCD-matrix contains two E2V CCD~44-82
detectors $2048\times4102$~pixels each, providing a $10\times10$~arcmin field
of view with a blind zone between the detectors. We observed four
overlapping fields with different exposures and under different seeing
conditions. Images of each of the fields were reduced separately. Data
reduction included bias subtraction
, correction
for non-linearity of each detector, flat-field correction,
removal of background air-glow emission and removing cosmic-ray hits.
{The final images were combined into a mosaic. Its central part is analysed
in this paper (shown in Fig.~\ref{fig:Ha_Hb_SII_OIII}c).}

We subtracted the underlying stellar continuum from optical images in \Ha\,
and \SII\ lines using the images obtained in the filters FN641 and FN712.
The residuals from the subtraction of the bright stars were masked in all
subsequent images in the paper. We did not observed the continuum around the
\OIII\ line, so in all figures containing \OIII\ image the underlying
continuum is not subtracted.

NIR images of the vdB~130 region were obtained with the ASTRONIRCAM infrared
camera \citep{Nadjip17} mounted in the  Nasmyth-1 focus of the 2.5-m
telescope. We observed the region of interest in four bands allowing us to
obtain the CO($v = 2-1$) ($\lambda$= 2.285 $\mu$m) emission line images and
the J, H, and K band photometric images. Observations were performed using
the dithering method, with the telescope shifting between individual frames
by $3-4$~arcsec. Each frame was corrected for non-linearity, dark current,
and flat field.

\begin{figure}
\includegraphics[width=\linewidth]{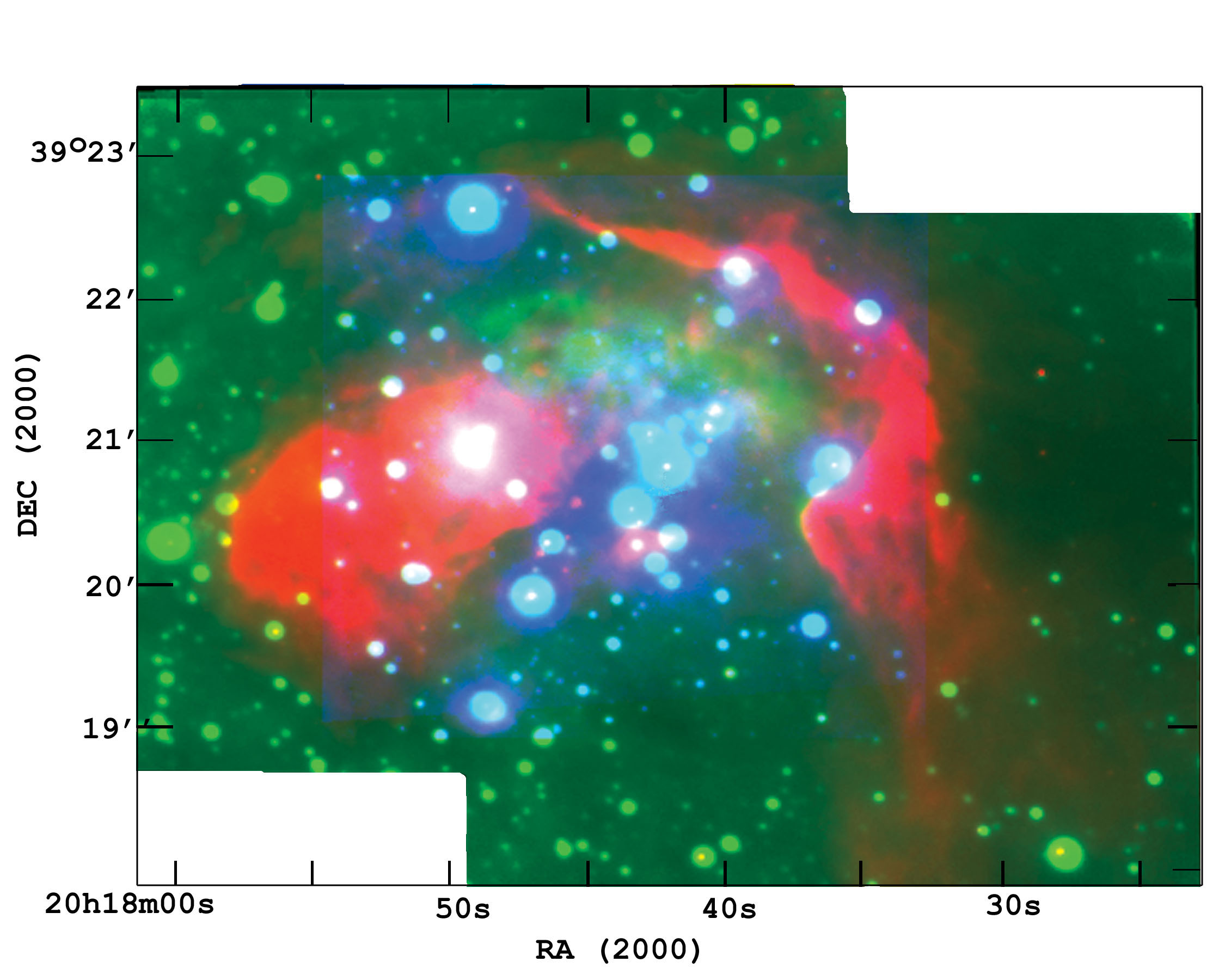}
\includegraphics[width=\linewidth]{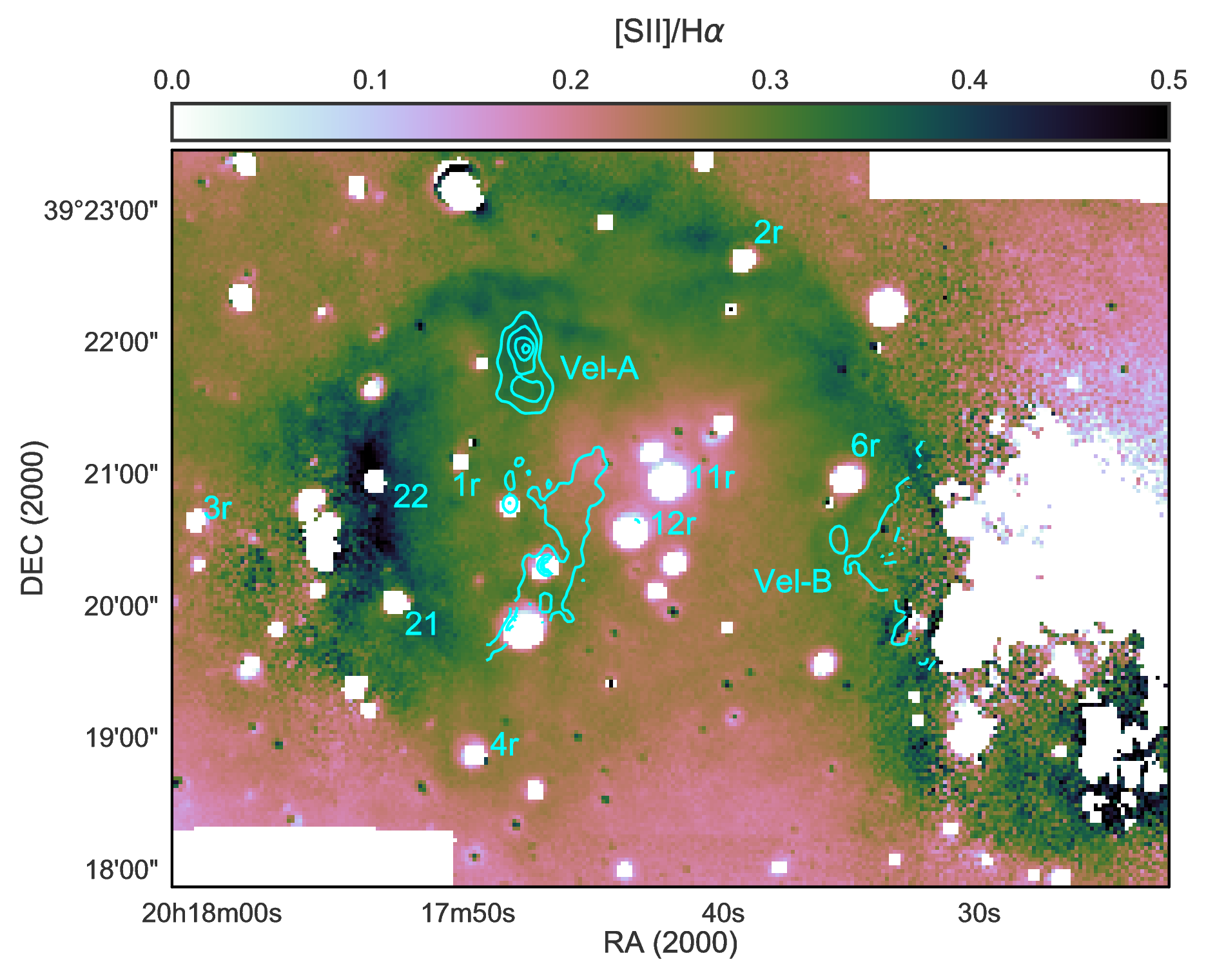}
\caption{Top: an RGB-image of the vdB~130 cluster area composed of the
8~$\mu$m band (red), \SII+\Ha\ line (green), and \OIII\ line (blue). Bottom:
a map of the \SII\, 6717, 6731/\Ha\ line intensity ratio. The areas of high
residuals after continuum subtraction around bright stars were masked.
Isocontours show the lines of the equal $V_{\mathrm{LSR}} = -7, -12, -17,
-22\ \kms$ in the \SII\ emission lines (see
Sec.~\ref{sec:discuss_A}. The areas of enhanced velocities Vel-A and Vel-B,
and the stars mentioned in the text are marked. Note that  \SII/\Ha\, ratio
is underestimated because the \Ha\, image is contaminated by \NII\,
emission.}  \label{fig:reg.A0}
\end{figure}

\subsection{Archival data used}

This work is partially based on observational data of \textit{Spitzer}
and \textit{Herschel} space telescopes. These data were downloaded from the
\textit{Spitzer} Heritage Archive\footnote{http://sha.ipac.caltech.edu} and the
Herschel Science
Archive\footnote{http://herschel.esac.esa.int/Science\_Archive.shtml}.

The studied region was observed by \textit{Spitzer} space telescope within
the framework of the program called `A \textit{Spitzer} Legacy Survey of the Cygnus-X
Complex'\footnote{http://www.cfa.harvard.edu/cygnusX} \citep{cygX}. The
images in the four bands at $3.6\,\mu$m, $4.5\,\mu$m, $5.8\,\mu$m, and
$8.0\,\mu$m were taken with the IRAC camera, and the $24\,\mu$m image was taken with
the MIPS  camera. The spatial resolution of the images taken with the two
cameras are equal to 0.6 and 2.45 arcsec per pixel, respectively.

We used \textit{Herschel} archival data obtained within the framework of the
`Hi-GAL: \textit{Herschel}  Infrared Galactic Plane
Survey'\footnote{https://tools.ssdc.asi.it/HiGAL.jsp} program
{\citep{Molinari10}}. This region around vdB~130 was imaged with the PACS
instrument at $70\,\mu$m and $160\,\mu$m and the SPIRE instrument at
$250\,\mu$m, $350\,\mu$m, and $500\,\mu$m with a spatial resolution ranging
from 3.2 to 14.0 arcsec per pixel. In our analysis we mostly use PACS
$160\,\mu$m and SPIRE $500\,\mu$m data with a resolution of 6 and 14 arcsec
per pixel, respectively.

\begin{figure*}
    \includegraphics[width=0.48\textwidth]{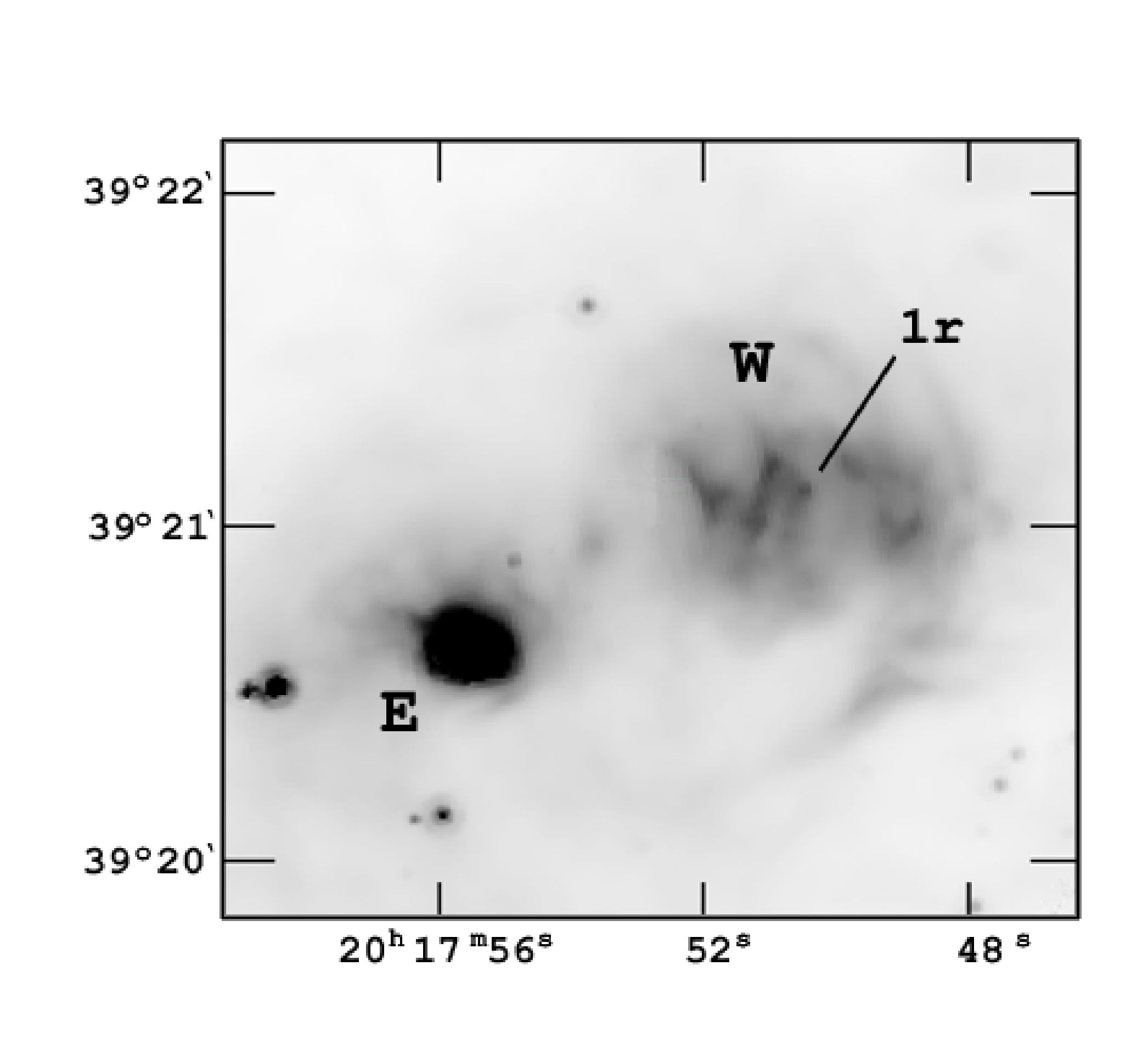}
    \includegraphics[width=0.48\textwidth]{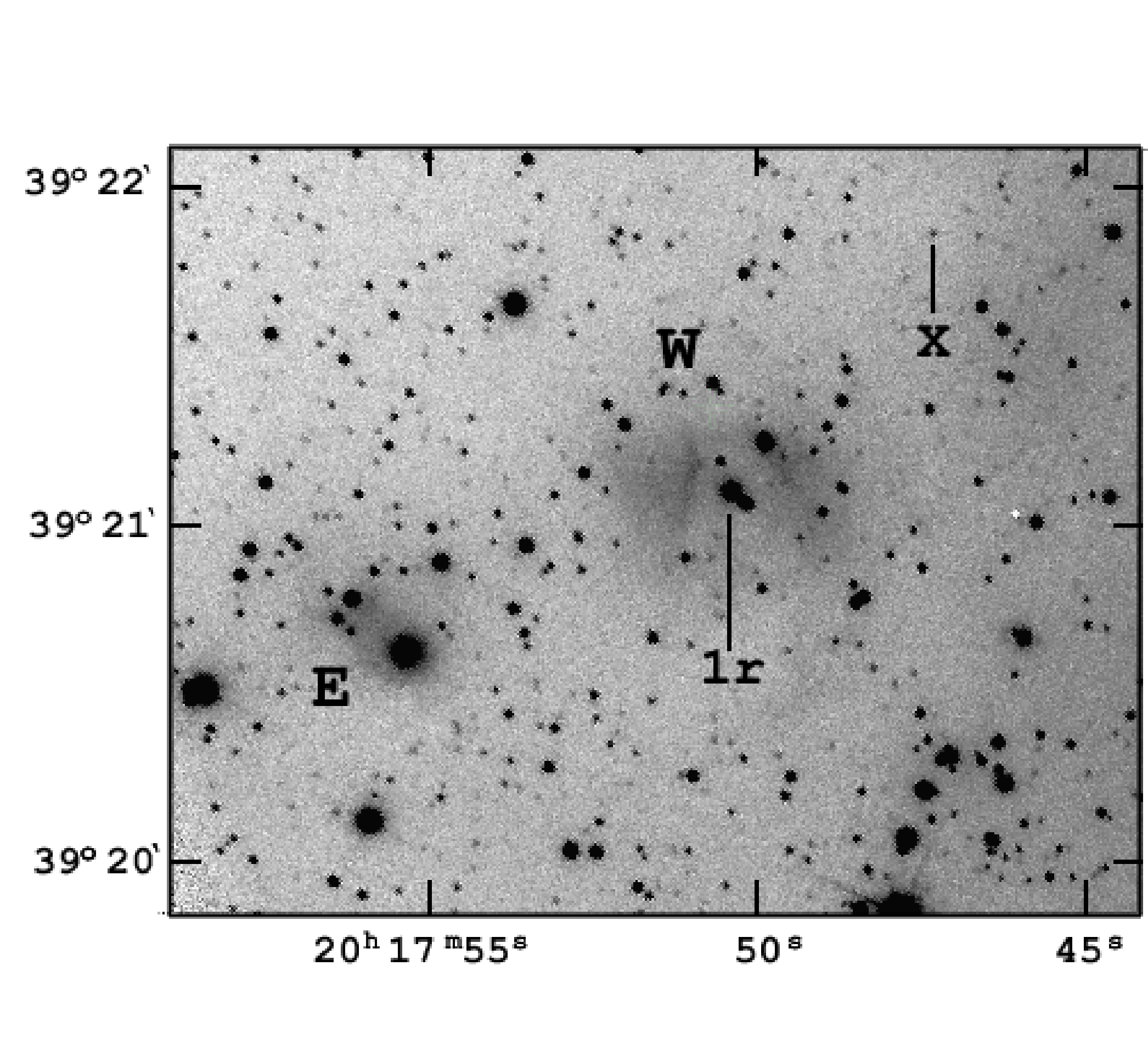}
    \caption{8~$\mu$m (left) and CO($v = 2-1$) (right) band images of Blob W and Blob E.
     The central star of Blob W, {1r (discussed in Section 3),} and the point source X are marked.}
    \label{fig:CO}
\end{figure*}

\section{Optical and IR emission line morphology of the region}

Fig.~\ref{fig:Ha_Hb_SII_OIII} shows the \SII\, 6717, 6731\AA, \Ha, \OIII\,
4959, 5007\AA\, and 8~$\mu$m band images of the studied region. As is evident
from the figure, \SII\, and \Ha\ line emission dominates in the north-western
part of the region, near the IR-shell, whereas \OIII\ emission is observed
primarily toward the cluster core stars 11r, 12r, and 13r as well as between
these stars and Blob W. Mid-IR emission, \SII+\Ha\ emission, and \OIII\
emission are further summarized in Fig. \ref{fig:reg.A0} (top). Three zones
stand out in this RGB-image. Red colour (IR emission) shows the locations of
the IR shell, Blob W and Blob E. Regions of bright \SII+\Ha\ emission are
indicated by green colour. Centred on the cluster core is the \OIII\ emission
shown with blue colour.

The overall structure of the area has been discussed in detail in
\citetalias{sit15} and is illustrated in Fig.~\ref{mapa}. Here we note some
peculiar features of this region. Among the peculiarities of the area, there
is a cavity with a size of approximately $0.6\times0.3$ arcmin
($0.3\times0.15$~pc), shown on insets of Fig.~\ref {fig:Ha_Hb_SII_OIII}. It
is distinguishable in all the observed lines. Also, several diffuse \OIII\
filaments are visible between \Ha\ Blob (around the vdB 130 core stars) and
Blob W. \OIII, \SII\, and \Ha\ line emission is noticeable in the direction
of Blob~W. The crescent-shaped filament in Blob~W, noticeable only in the IR,
neighbours the \OIII\ emission region in the plane of the sky (Fig.
\ref{fig:Ha_Hb_SII_OIII}d and \ref{fig:reg.A0}, top). An \OIII\ emission
brightening is evident to the west of Blob W (at the base of the
crescent-shaped filament).

In \citetalias{sit15}, signatures of a weak shock were detected between
Blob W and Blob E (Fig. \ref{fig:reg.A0}, bottom), based on the distribution
of the $I($\SII\,$6717,6731)/I($\Ha$)> 0.4$ ratio {\citep{allen}}. The
shock front coincides with the faint \SII\ filament at the eastern part of
the region (Fig. \ref {fig:Ha_Hb_SII_OIII}a). Signs of a shock wave are also
visible around and inside the cavity (Fig.~\ref{fig:reg.A0}, bottom). The
shock front to the north of the cavity looks like an extension of the shock
front located between Blob W and Blob E. Note  that, according to the
analysis reported in \citetalias{sit15}, only the western part of the
X-shaped structure with an increased $I($\SII\, $6717,6731)/I($\Ha$)$ ratio
can be considered as an indication of a shock front.

Blob W is located in the area affected by UV radiation and stellar winds from
the cluster core. Traces of this influence are clearly seen in Fig.
\ref{fig:CO} (left). The brightest 8, 5.8 and 3.6~$\mu$m band emission
features in Blob W constitute a system of filaments visible in polycyclic
aromatic hydrocarbon (PAH) emission and oriented approximately in the
north-south direction. As is the case with ionized gas regions, PAH emission
is dominant in the outer regions of the blobs and characterizes the blob
shell, whereas hot dust emits at 24~$\mu$m, primarily in the central parts
of Blob W and Blob E (Fig. \ref {fig:vdB130}, bottom).

The cluster star 1r (B1V) \citep{rac68, tatar16} and a Class~I protostar (see Section 6) are located
inside Blob W, `destroying' it from inside (Fig. \ref{fig:CO}, right). The
structure of Blob W is also clearly seen in the CO($v = 2-1$)
$\lambda$2.285~$\mu$m band image of the region. A central cavity around the
star is bounded from east and west by thin filaments surrounded by more
diffuse CO emission, and opened from north and south. The CO filaments appear
to border the 24~$\mu$m emission and are spatially coincident with the
8~$\mu$m filaments, possibly, representing the broken shell of the bubble.
The CO $\lambda$2.285~$\mu$m band emission brightness depends on the medium
density, and amplifies in a shock wave \citep{draine84}. In our case, both
conditions are satisfied: there are traces of a shock wave, and, according to
the estimates of \citet{tatar16}, the density inside the blob is quite high
as the extinction there amounts to about 10~mag.

\section{Kinematics of ionized gas}

\begin{figure}
    \includegraphics[width=\linewidth]{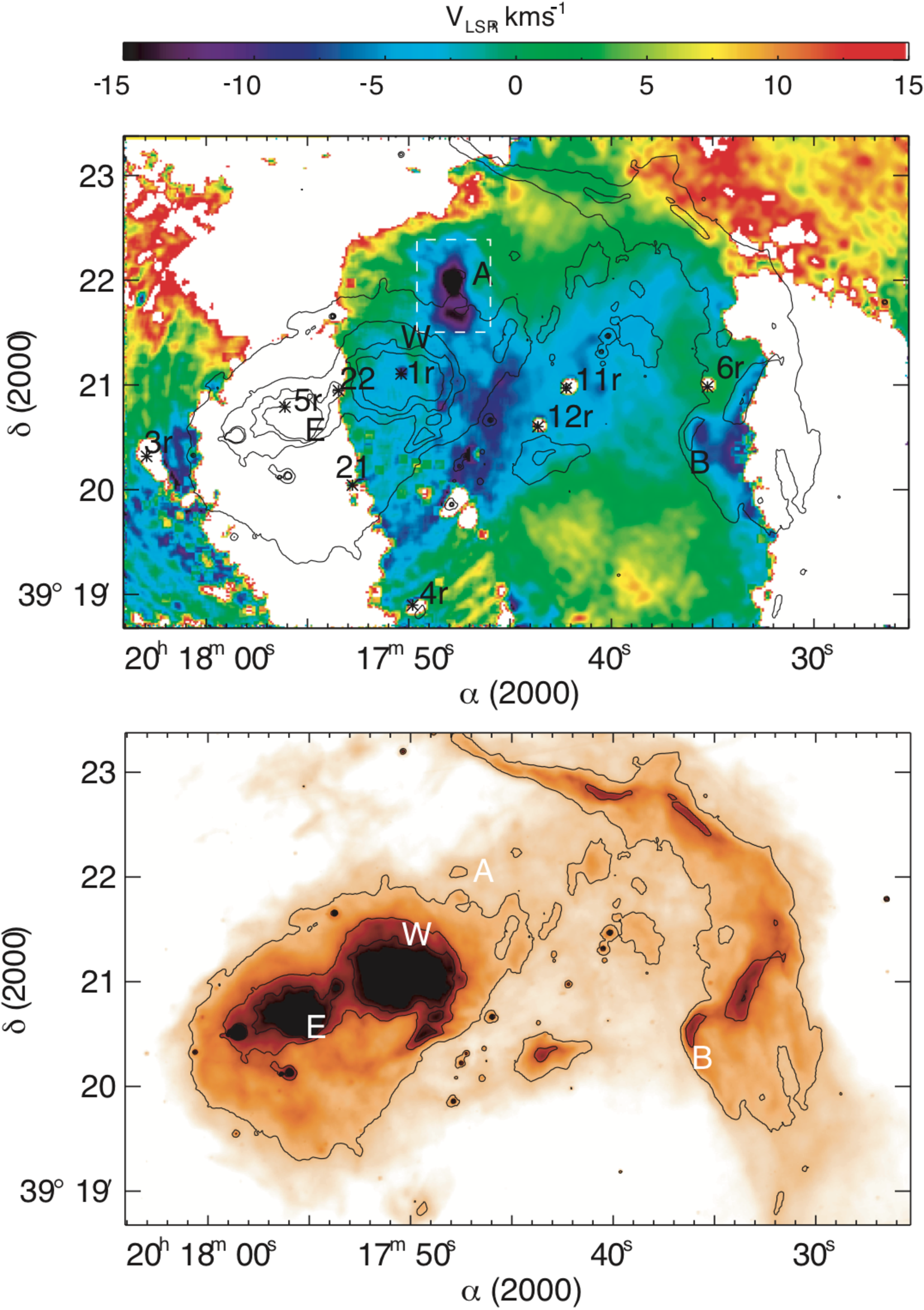}
    \caption{The \SII\, 6717, 6731 \AA\, line-of-sight velocity $V_{\mathrm{LSR}}$ field in the vdB~130
    cluster area with overlaid $8\,\mu$m band isophotes (top).
        The 8~$\mu$m band image of the region according to \textit{Spitzer} data (bottom). The numbers
        indicate the vdB~130 stars, whereas letters A and B mark the high-velocity areas. The shock front
        passes through the eastern dust cloud boundary (through stars 4r, 21 and 22). All regions with
        $S/N < 3$ were masked.
        For a zoomed image of Vel-A region (inside the dashed rectangle)
        see Fig.~\ref{fig:reg_A}.}\label{fig:velocity}
\end{figure}

In the top panel of Fig.~\ref{fig:velocity} we show the distribution of
line-of-sight velocities, $V_{\mathrm{LSR}}$, in the vicinity of vdB~130
obtained by fitting two single-component Voigt profiles to the observed
\SII\, 6717,6731 \AA\, line profiles in each pixel of the data cube. For
comparison, the bottom panel of Fig.~\ref{fig:velocity} demonstrates the IR
image of the same region in the 8~$\mu$m band. The region under investigation
is bounded by the shock front from the east (from the side of the Cyg~OB1
association) and by the IR shell enveloping the north-western ionization
front \citepalias{sit15}.

An analysis of the velocity field together with the line profiles
decomposition revealed the following (Figs \ref{fig:velocity} and
\ref{fig:profile}).

\begin{enumerate}
    \item  In general, the \SII\ line profiles are symmetric, but in several locations
    they exhibit broadening or asymmetry and might be decomposed into two components with
    a velocity separation of up to $40 \kms$ (see Fig.~\ref{fig:profile}, profiles \#2, \#3, \#6, \#10, \#12).

    \item The $V_{\mathrm{LSR}}$ range of the main component in the area is $\sim-1...+6\kms$
    (green areas in Fig. \ref {fig:velocity}, top). These values agree with the velocity estimates for the
    ionized hydrogen and the molecular cloud \citepalias[see][and references therein]{sit15}. The main
    velocity component of \Ha\ and CO lines, observed outside the cluster in the direction of the
    molecular cloud, corresponds to $V_{\mathrm{LSR}}\sim3...5\kms$. (In this
    direction galactic rotation velocities are positive up to a distance of 5~kpc.)

    \item  In the vicinity of the B-stars 11r and 12r (in \Ha\ Blob), velocities of
    $V_{\mathrm{LSR}}\sim -7...-3 \kms$ are dominant.

    \item $V_{\mathrm{LSR}}\sim-3\kms$ is observed in Blob W. However, {as} $\mathrm{S/N} < 3$ for most of the \SII\ line
    observations within the blob, accurate velocity measurements are impossible.
    Ionized gas emits at velocities of
    $V_{\mathrm{LSR}}\sim -7 ... -5\kms$ to the south-west of Blob W (outside the crescent-shaped IR filament).
    The same velocities, $V_{\mathrm{LSR}}\sim -7 ... -3\kms$, are observed in the direction of some parts
    of the crescent-shaped filament itself.

    \item  Two regions are distinguished that have high negative radial velocities: region Vel-A to the north-west
    of Blob W,
    and region Vel-B in the IR shell (see Fig. \ref{fig:velocity}, top). A broadened \SII\ line profile in region
    Vel-A can be decomposed into two components in some directions (Fig. \ref{fig:profile}, profiles \#2, \#3).

\end{enumerate}

An outline of region~Vel-A in Fig. \ref {fig:reg_A}a is eight-shaped. The
plotted position-velocity (PV) diagrams for this region clearly reveal an
approaching part of the velocity ellipse (diagram \#1 in Fig.~\ref{fig:pv}).
The mean ionized gas velocity $V_{\mathrm{LSR}}\sim -30...-25 \kms$ in the
northern part of the `eight', which coincides with the cavity, and
$V_{\mathrm{LSR}}\sim -15 \kms$ in the southern wall of the cavity. The size
of the high-velocity area is $0.7 \times 0.2$~arcmin ($0.35\times0.1$~pc).
The velocity dispersion in region Vel-A is equal to $\sim 25 \kms$, which is
far greater than anywhere inside the IR shell.

Region Vel-B is located in the direction of the IR shell at velocities in the
range of $V_{\mathrm{LSR}}\sim -10 ... +1\kms$ (see top panel in
Fig.~\ref{fig:velocity} and 145--148~arcsec positions on PV-diagram \#2 in
Fig. \ref {fig:pv}). Region Vel-B is coincident with a thin filament emitting
in the IR and all the considered visual band lines.

Thus, in the vicinity of the cluster, we see variations in the line-of-sight
velocity of ionized sulphur from $-30 \kms$ to $5 \kms$. Regions with
dominant $V_{\mathrm{LSR}}<-2\kms$ motions (shown by cyan-blue in Fig.
\ref{fig:velocity}a) generally coincide with the \OIII\ emission areas
(Fig.\ref{fig:Ha_Hb_SII_OIII}c). As follows from PV-diagram \#3 in Fig.
\ref{fig:pv}, the gas inside the IR shell around stars 11r and 12r expands
with a velocity of up to $15 \kms$ (Fig.~\ref{fig:vdB130}~bottom, see also
Fig. 12d in \citetalias{sit15}). This conclusion is consistent with the
assumption on the expansion of the \Ha\ Blob and the IR shell around vdB~130
core B-stars and its interaction with Blob W (see \citetalias{sit15}). In
some directions (Fig.~\ref{fig:profile}, profiles \#6, \#12) we observe
two-component profiles that may correspond the approaching and the receding
walls of the shell. Presumably, it is the influence of the wind and radiation
of the cluster core stars that is responsible for the formation of the shock
between the blobs.

Note that the large scatter of the main component radial velocities $V_{\rm
LSR} \sim -1... +6 \kms$ may hamper distinguishing between local low-velocity
motions (like the expansion of \Ha\ Blob) and those on a large scale,
determined by galactic rotation.

\section{On the origin of the shock waves in region Vel-A}
\label{sec:discuss_A}

The vdB~130 cluster neighbourhood is a compact region of ongoing star formation in
the supershell around the Cyg OB1 association, populated by young stars, which actively
interact with the surrounding inhomogeneous interstellar medium. Tracing individual
relations between various components of the complex is fairly hard. We therefore consider
different possible explanations for the nature of the detected high-velocity motions of ionized gas.

\subsection{What do we know about region Vel-A}

The region is distinguished by the highest negative radial velocities (Fig.
\ref{fig:reg_A}, top left) over the entire vdB 130 cluster area. The
following additional peculiarities are observed in the direction of
region~Vel-A:

\begin{figure*}
    \includegraphics[width=\linewidth]{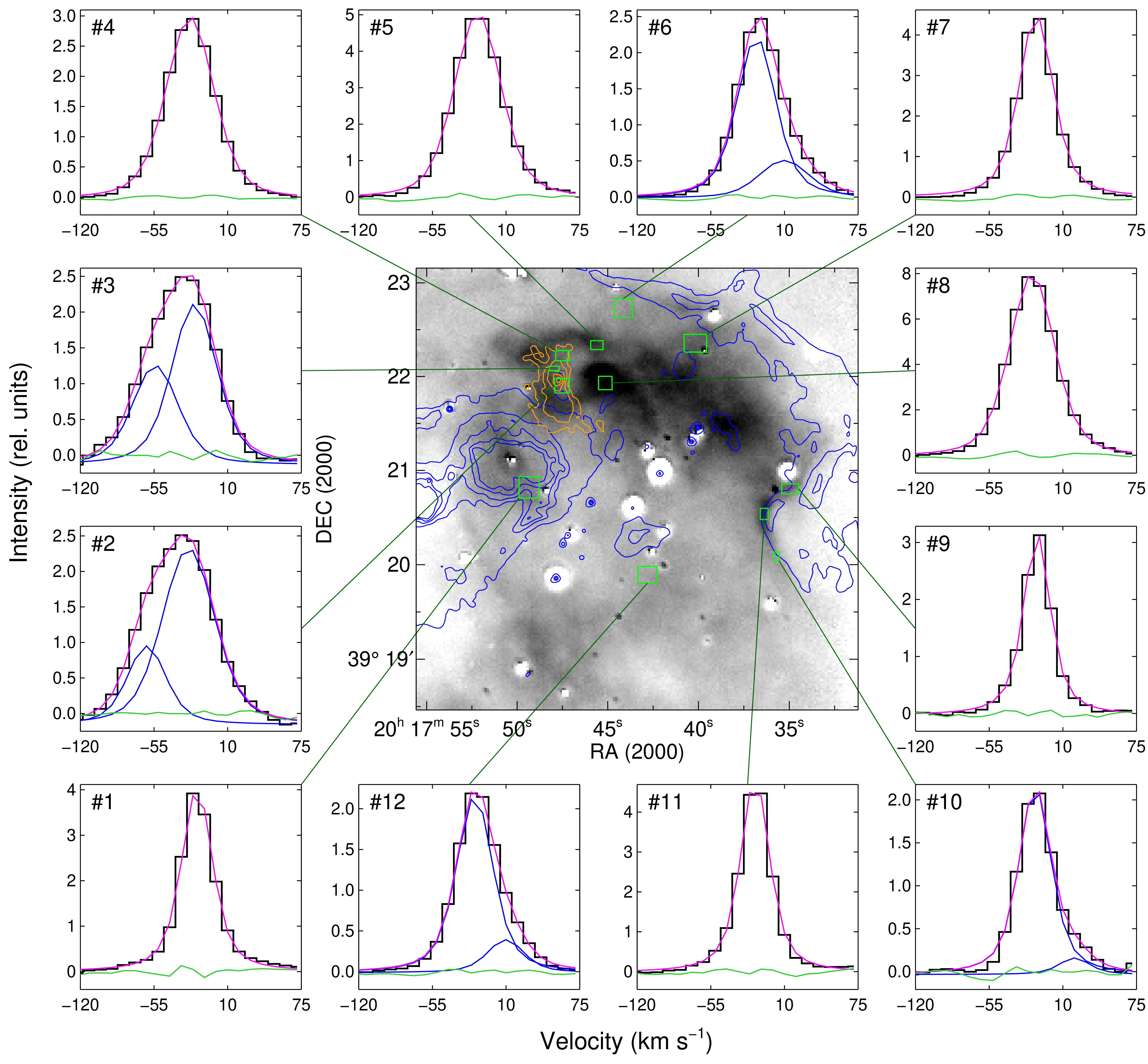}
    \caption{Localization of the \SII\ (6731~\AA) line profiles in the \SII\, line image of the region, and examples
    of profiles and their decomposition. Black colour shows the observed line profiles, blue colour shows individual
    decomposition components, magenta colour represents the final model, and model subtraction residuals are plotted in green.
    The areas where the corresponding profiles were averaged are also shown in green on the image. Blue contours overlaid on the image
    show the distribution of emission in 8~$\mu$m band. Orange contours show the lines of equal
    $V_{\mathrm{LSR}} = -5, -10, -15, -20, -25\ \kms$ in the [S~\textsc{ii}] emission lines for the region Vel-A}\label{fig:profile}
\end{figure*}

\begin{enumerate}

    \item Two IR spots with the sizes of approximately 0.1 and 0.2~arcmin
(0.05 and 0.1~pc) are projected onto lobes
 of the `eight' and are visible in all \textit{Spitzer} bands as well as in JHK bands. (These spots are designated
 as `1' and `2' in  Fig. \ref{fig:Ha_Hb_SII_OIII}d and  Fig.~\ref{fig:reg_A} (top right).)

\item The radial velocity distribution in the direction of the `eight' is related to the location of the IR spots (Figs \ref{fig:Ha_Hb_SII_OIII}d and
 \ref{fig:reg_A}ab).

\item Region Vel-A and the IR spots are projected onto a $0.3\times0.15$~pc
sized cavity in the distribution of optical emission (see Section 3), (Figs
\ref{fig:Ha_Hb_SII_OIII}abc and \ref{fig:reg_A}).

\item Ionized sulphur is observed to have supersonic velocities in the centre and on the northern border
of the cavity (Fig. \ref{fig:reg_A}); traces of a faint shock wave are also present (Fig.~\ref{fig:reg.A0},
see also \citetalias{sit15}).
\end{enumerate}

\begin{figure*}
\includegraphics[width=0.47\linewidth]{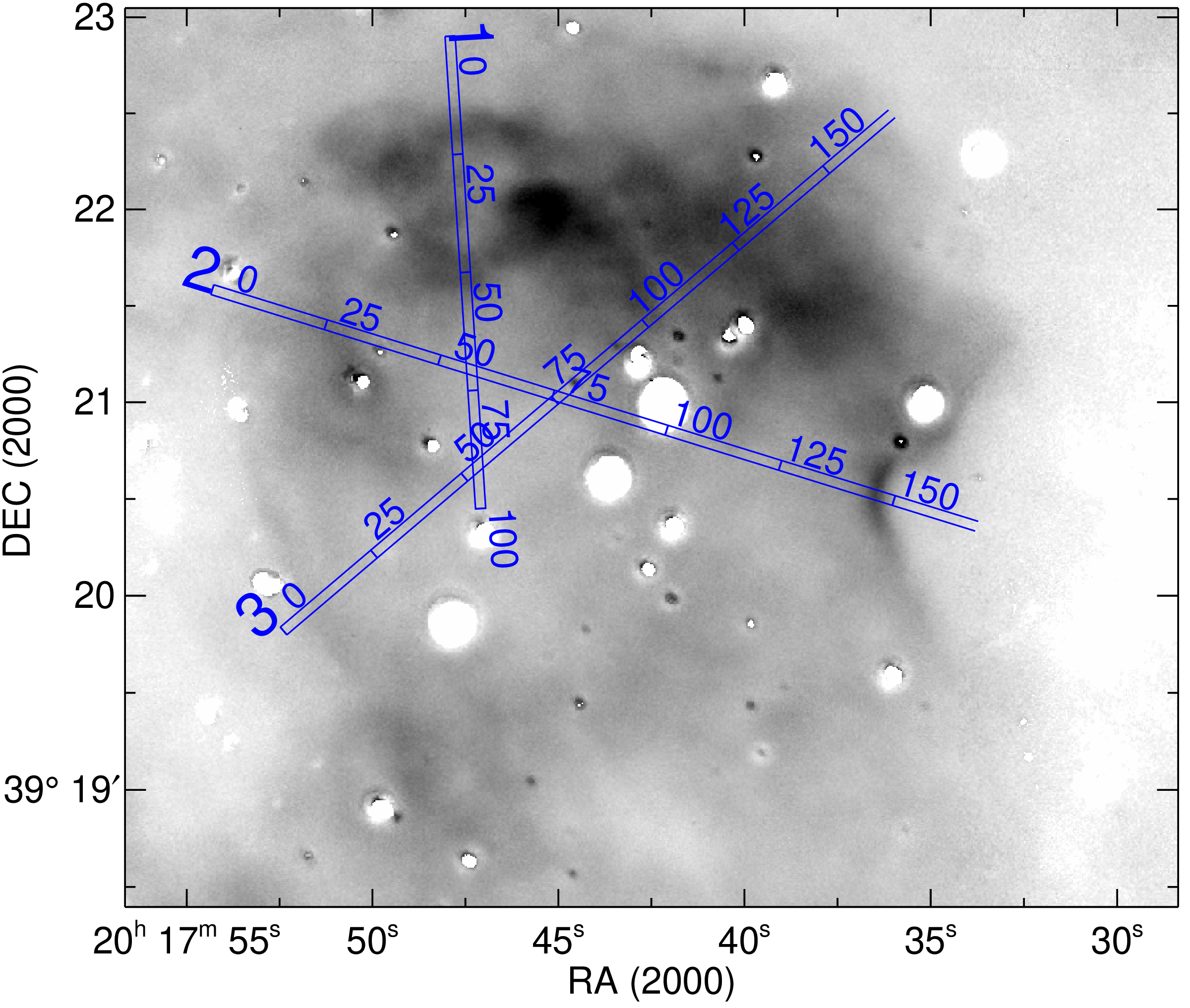}~\includegraphics[width=0.51\linewidth]{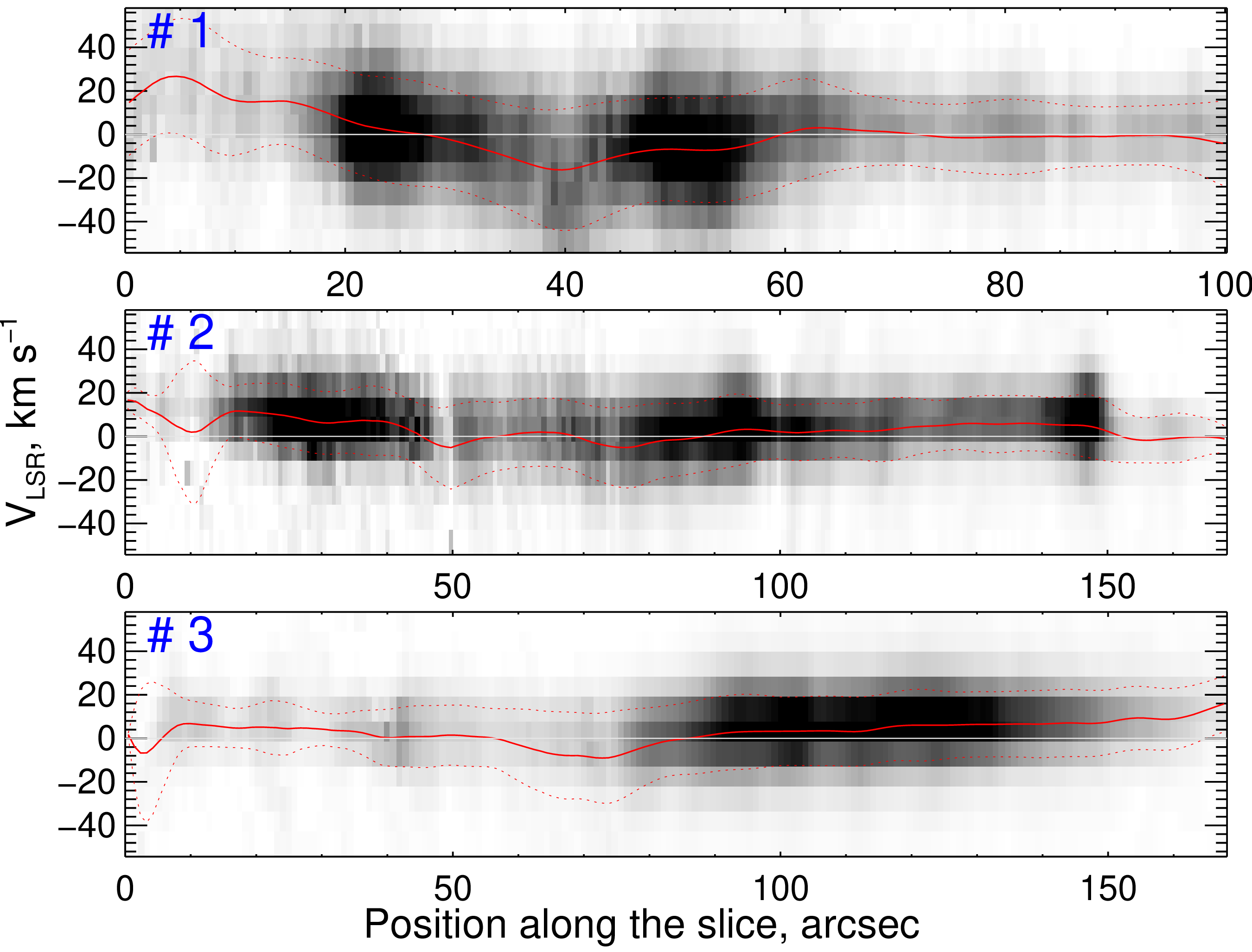}
\caption{Left panel: position of PV-diagrams on the \SII\ image. The
corresponding \SII\ 6731 line PV-diagrams are shown in the right-hand panels.
The red solid curves show the median velocity distribution, and the dotted
curves represent its standard deviation.}\label{fig:pv}
\end{figure*}

\begin{figure}
\includegraphics[width=\linewidth]{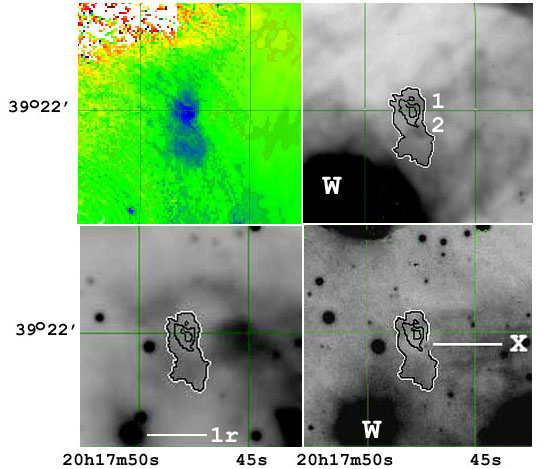}
\caption{Top left:Velocity distribution in region Vel-A (top left). The blue
colour corresponds to radial velocities less than $-10 \kms$. Top right:
Radial velocity $V_{\mathrm{LSR}}\leq-10 \kms$ isocontours, combined with the
8~$\mu$m band image of the region. Bottom left: \Ha\ image of the region.
Bottom right: \OIII\ image of the region. The blob W, the cluster star 1r and
the object X are indicated. Bright IR features are marked by numbers.}
\label{fig:reg_A}
\end{figure}

We have detected neither any point sources (Class I-III protostars), nor any
stars brighter than $V\simeq22$ mag (or K~$\simeq18$ mag) inside the region
Vel-A. A sole exception is a point source X ($\alpha = 20^{h}17^{m}47.34^{s},
\delta=39^{\circ}21^{'}53.7^{''}$) located near the western edge of the
southern lobe (Figs \ref {fig:Ha_Hb_SII_OIII}{c} and \ref {fig:CO}).
In the optical band, this source stands out only in  \OIII\ images and in JHK
bands; it is also visible at the limit of sensitivity in 3.6 and 4.5~$\mu$m
\textit{Spitzer} bands. The source X is most probably a star with a chance
projection of a foreground dust clump. This star is therefore observed
through a layer of absorbing material. However, we were not able to determine
whether or not it belongs to the cluster.

\begin{figure}
\includegraphics[width=\linewidth]{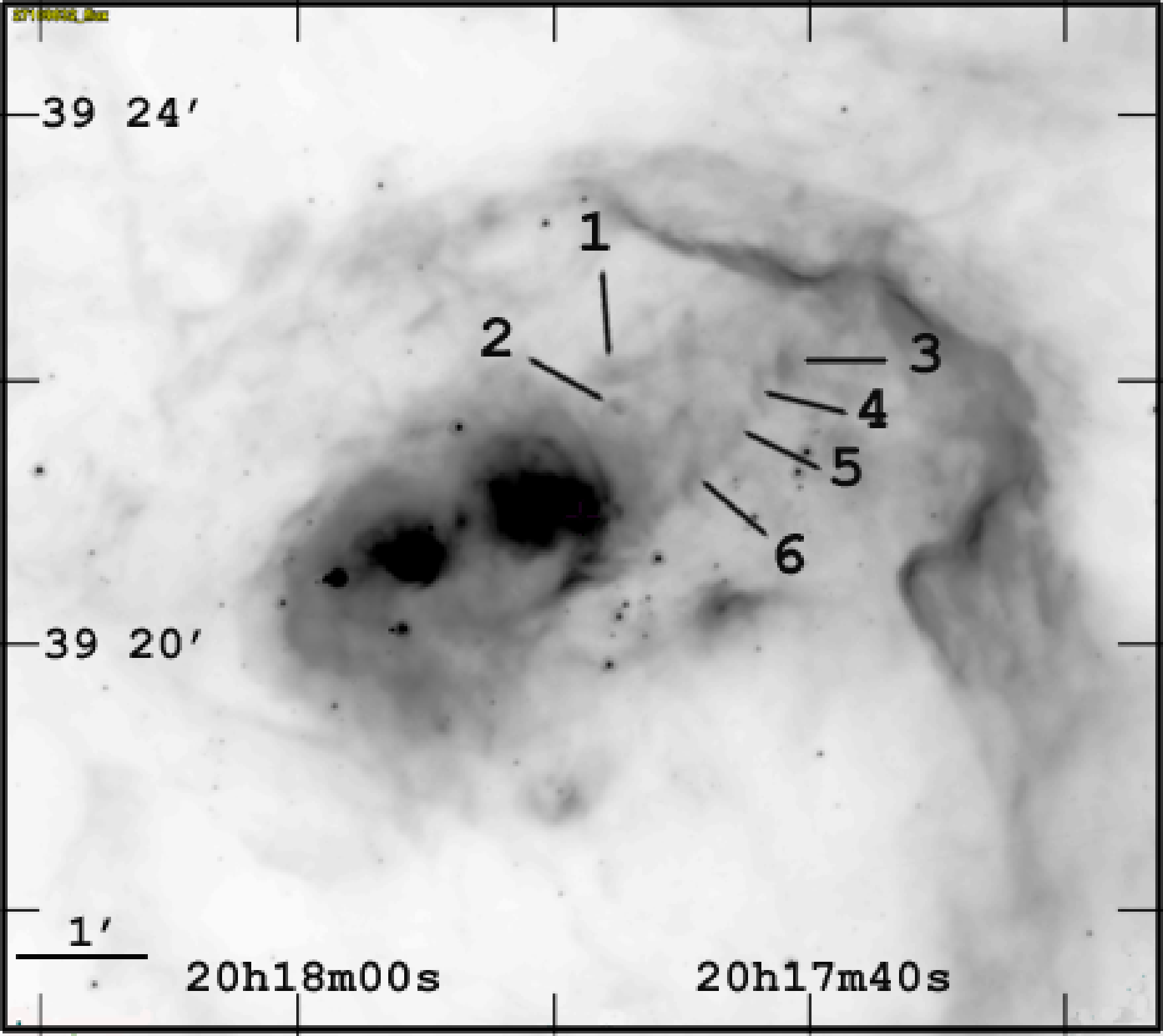}
\includegraphics[width=0.74\linewidth, angle=270]{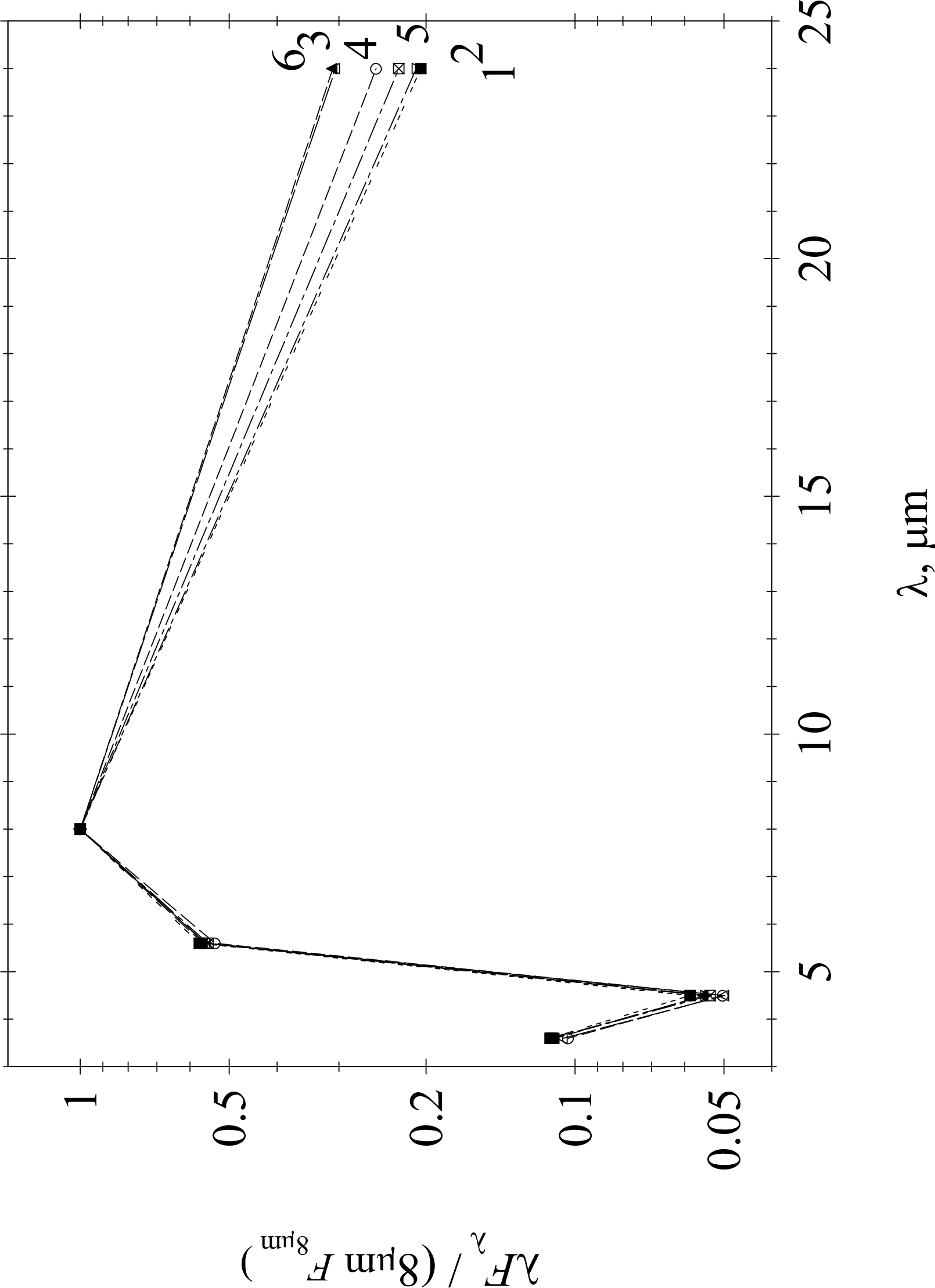}
\caption{Top: 8~$\mu$m band image of the vicinity of the cluster vdB 130,
some bright IR features are marked by numbers. Bottom: SEDs of the marked IR
emission spots. {All the curves are normalized to spectral density at
a wavelength of $8\,\mu$m}.} \label{fig:SEDs}
\end{figure}

Considering the fact that the two IR emission spots (Fig. \ref{fig:reg_A},
top right) are projected onto the region Vel-A, we searched for possible
embedded sources in the dust clumps. We plotted the spectral energy
distributions (SEDs) for spots 1 and 2 and also for four other IR emission
spots located along the northern boundary of the molecular cloud (Fig.
\ref{fig:SEDs}, top) as well as for several less bright areas in the vicinity
of these spots. The surface brightness of these spots in the range 3.6~$\mu$m
-- 8~$\mu$m  was measured with a $\sim7$ arcsec aperture. For the 24~$\mu$m
band, $\sim$10 arcsec aperture was used. The size of the aperture was chosen
so that the visible boundaries of the spots were outside it. The background
was determined outside the IR shell. The [3.6] -- [4.5] and [4.5] -- [5.8]
colours were estimated using the normalizing data from \citet{gut08}.

Fig. \ref{fig:SEDs}, bottom, shows the SEDs of all the spots investigated. { The
curves are normalized to spectral density at a wavelength of 8~$\mu$m. The SEDs in all
the marked areas have similar shapes}. The
deviation of the points from the average value in the 3.5 -- 8~$\mu$m range
does not exceed 10\%. The accuracy decreases in the 24~$\mu$m band due to the
smaller angular resolution. { Such a similarity of the SEDs for different spots
and also for the common IR shell may hint at the similarity of physical conditions or, specifically,
may be an indication of the absence of bright embedded objects. However, we should note that
we have only constructed SEDs for $\lambda<25\,\mu$m that may not be sufficient to detect
deeply embedded sources.}

Since the shapes of the SEDs of all the investigated IR spots look alike, we can estimate the typical
values of the corresponding colour indices, $[3.6] - [4.5] = -0.05$~mag, $[4.5] - [5.8] = 3.2$~mag. Correction of the
colour indices for interstellar extinction using $E(B-V)\sim1$ \citep{tatar16} yields $[3.6] - [4.5]\sim -0.15$~mag
and $[4.5] - [5.8] \sim 3.15$~mag. According to the colour index diagram presented by \cite{ybarra14}
(see their Fig. 1), colours of all the shown IR spots are consistent with those of PDRs with a cloud thickness
of $A_{v} = 4 - 5$~mag in front of the far-UV source. This fact also agrees with the assumption of the ongoing star formation in the region.

\subsection{Possible origin of regions Vel-A and Vel-B}

\subsubsection{Region Vel-A as a result of the influence of a wind from a young star onto the ISM}

Considering the fact that the high-velocity $0.35\times0.1$~pc sized region
Vel-A is partly projected onto a cavity with a size of $0.3\times0.15$~pc
(Fig. \ref{fig:reg_A}), the most likely cause of the supersonic motions there
can be attributed to stellar wind. Traces of a faint shock wave have indeed been
detected inside this cavity and in the northern part of its shell (Fig.~\ref
{fig:reg.A0}, bottom, see also \citetalias{sit15}), which agrees with the
supersonic velocities of ionized sulphur found here independently:
$V_{\mathrm{LSR}}\sim -30\kms$ inside the cavity (between IR spots 1 and 2)
and $V_{\mathrm{LSR}}\sim -10...-15\kms$ at the outskirts (see 20--50~arcsec
positions on PV-diagram \#1 in Fig. \ref {fig:pv}).

The observed average velocity gradient towards the centre of the cavity, the
velocity ellipse, and a confident decomposition into two components allow us
to conclude that the peculiarity of region Vel-A may be related to the
expanding local shell of ionized gas, albeit the source of the radiation
and stellar wind is unclear. The only point source inside the cavity is the
source X mentioned above. We do not know, however, if this source is a star
belonging to the cluster. According to the DSS catalog and our own optical
images, there are no other point sources in the cavity brighter than 23~mag.
One should not rule out, though, that in this area with high and
inhomogeneous extinction the source of stellar wind may be hidden by a
{ compact gas-dust knot} \citep{rac68, tatar16}. For example, $A_{v}\sim
12$ mag would be sufficient for a B2-star located at the distance of the
cluster to have $m_{v}\leq 23$ mag.

The radial velocity distribution in the southern part of the `eight' has some
interesting features (Fig. \ref {fig:reg_A}, top left). Smaller radial
velocities (grey-green details) are observed in the direction of the centre
of IR spot 2, compared to those at the outskirts of the `eight' (blue
details). Such a radial velocity distribution suggests that a hypothetical
star is located behind a dust clump, where the shock wave propagates slower
due to a higher density. In other words, the southern IR feature (spot 2) may
be a line-of-sight cloud located in front of the point source, but {still} in the
vicinity of vdB~130 (top right panel of Fig. \ref{fig:reg_A}).

Possibly, for that same reason of high and inhomogeneous interstellar matter
density in the region  we see only the wall of the shell around the cavity,
approaching us at a supersonic speed. Note also that the `optical' cavity is
partly filled with dust clumps, namely IR spots 1 and 2, although we cannot
rule out that these spots are located in the front wall of the expanding
cavity shell.

As for the Vel-B region, it is in an expanding shell. Perhaps here we observe
the ionized gas flow around a small protrusion of part of the PDR of the
western rim of the shell (Fig.~\ref{fig:reg.A0},top). As a result, the
line-of-sight projection of the shell expansion velocity increases. And we
see both negative and positive components of the velocity (see
Fig. \ref{fig:profile}, spectrum \#10, and Fig. \ref{fig:pv}, PV diagram \#2).

\subsubsection{Region Vel-A as a result of an outflow from a protostellar disc}

{While a wind hypothesis seems to be a preferable one, we cannot completely ruled out that
the observed line-of sight velocities in region Vel-A (Fig. \ref
{fig:reg_A}a) may have arisen due to a bipolar outflow} from
the disc of a Class 0 protostar \citep[see, e.g.][]{frank14}. In our case,
both the northern and southern parts of region Vel-A demonstrate negative
average line-of sight velocities ($-30$ and $-15 \kms$), and the
$V_{\mathrm{LSR}}$ of the main gas component due to the galactic rotation is
equal to $-1...+6 \kms$. Such a line-of sight velocity distribution can indeed be
attributed to bipolar outflow, if the disc is tilted by a small angle to the
plane of the sky, and the outflowing gas propagates in a solid angle whose
projection onto the plane is more than twice the tilt angle.

However, bipolar outflows from discs of Class 0 protostars are observed in
molecular SiO and CO lines, as well as those of atomic oxygen
\citep{frank14}. In our case, the supersonic velocity region is seen in the
ionized sulphur lines (the maximum line-of-sight velocity relative to the
main component is $V_{\mathrm{LSR}}\sim -40 \kms$, see decomposition of
\SII\, line profiles \#2 and \#3 in Fig.~\ref{fig:profile}.). One might
suggest that the ejected `cold' material, moving at supersonic speeds, could
have heated to sulphur ionization temperatures due to collisions with dense
interstellar clouds. Also, region~Vel-A is located within the range of
influence of UV radiation and winds from the vdB~130 cluster stars, so that
we cannot exclude their contribution to the outflow ionization \citep[see
also][]{mcleod2015}.

\begin{figure}
\includegraphics[width=\linewidth]{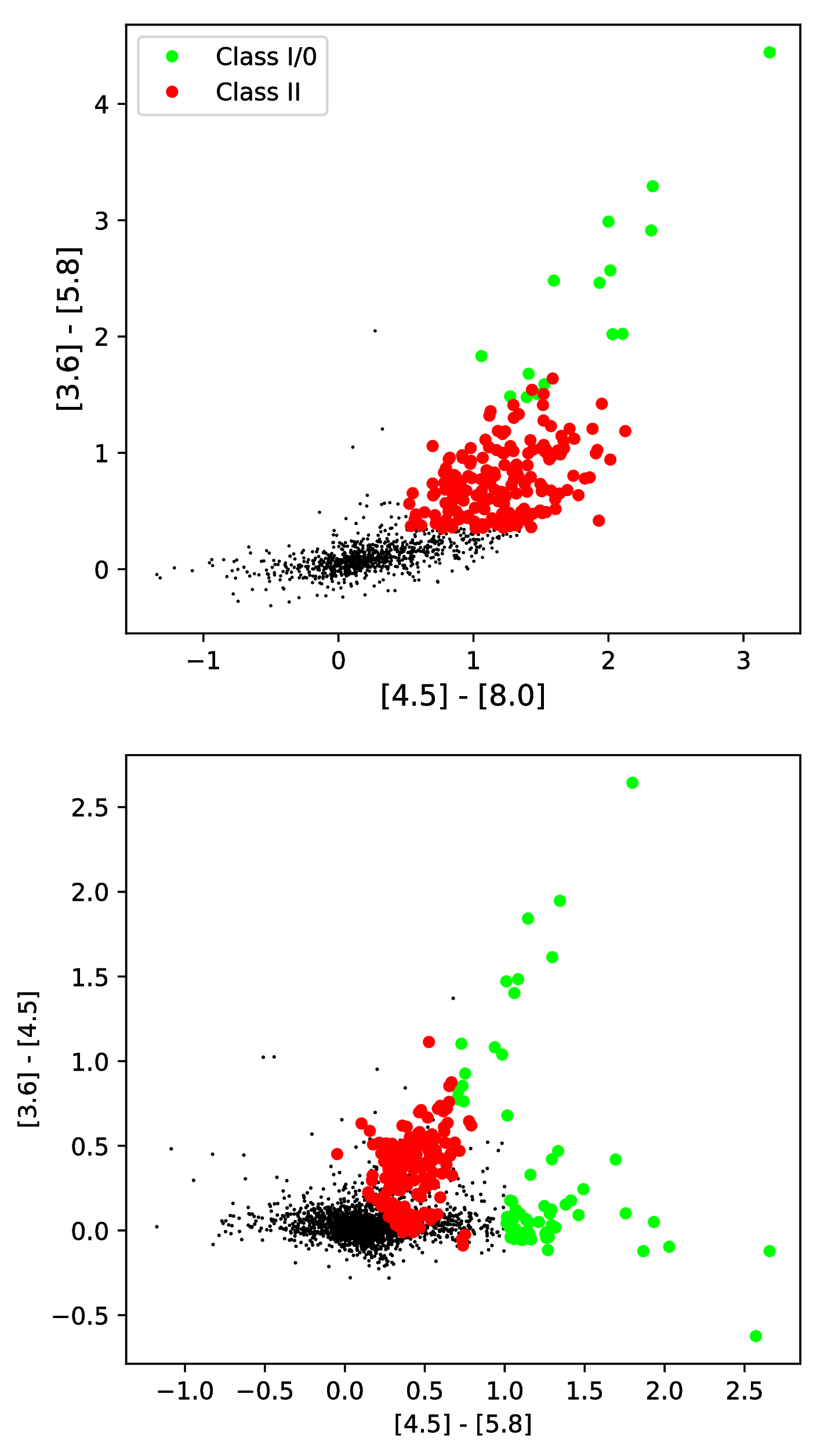}
\caption{The colour-colour diagrams used to separate Class I objects from those
belonging to Class II, according to the method of \citet{gut08}. Class I
sources are shown by green, and Class II sources are marked by red.}
\label{fig:YSOsel}\end{figure}

\begin{figure}
\includegraphics[width=\linewidth]{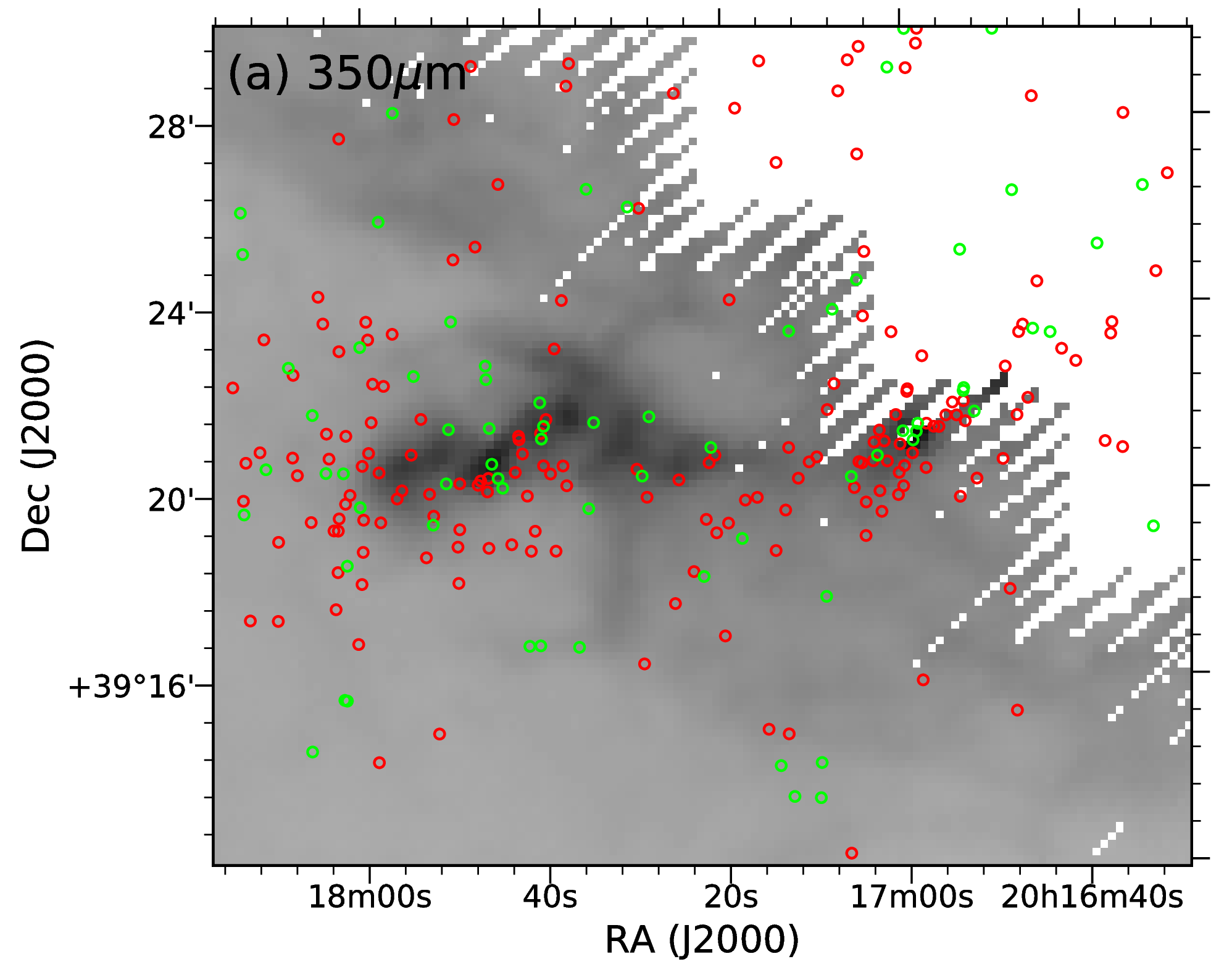}

\includegraphics[width=\linewidth]{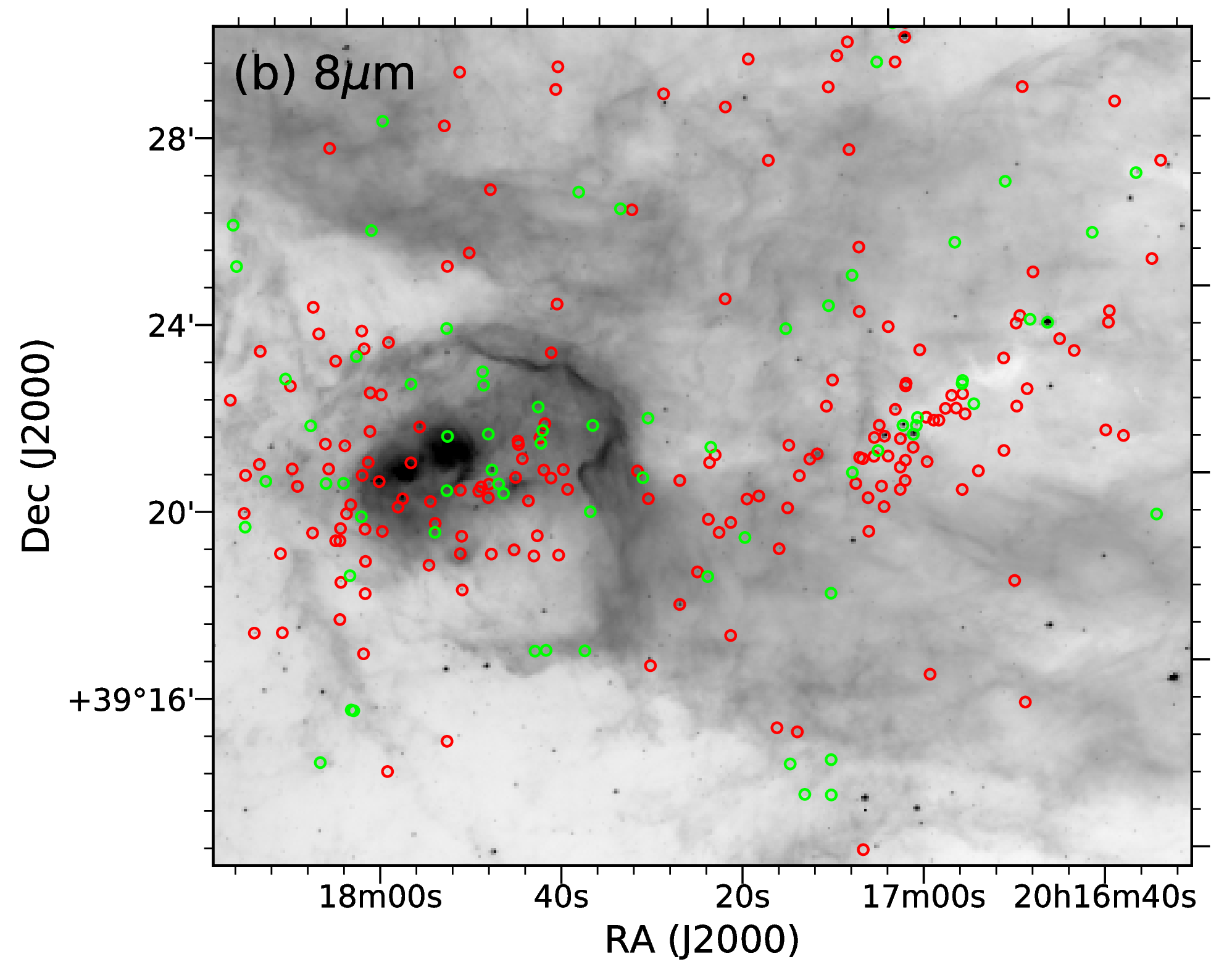}

\includegraphics[width=\linewidth]{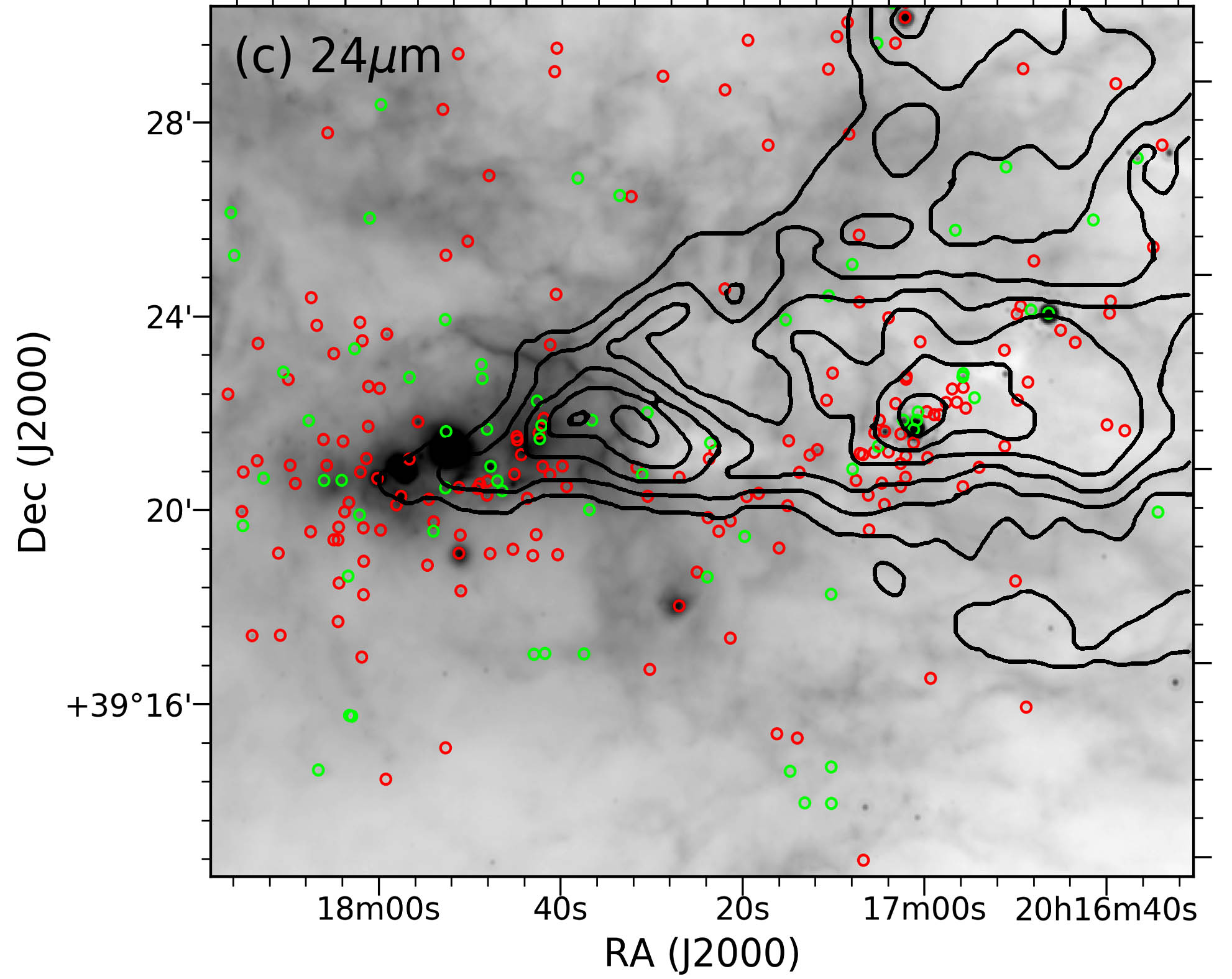}
\caption{Positions of the selected YSOs in the $350 \mu$m (a), $8 \mu$m  (b)
and $24 \mu$m (c) band images of the area based on the data of the
\textit{Herschel} and \textit{Spitzer} space missions. Isocontours of the
molecular cloud are presented in panel {\it c} \citep{sch07}. The green colour
shows the objects belonging to Class I, whereas Class II sources are shown in
red.} \label{fig:YSOloc}
\end{figure}

\begin{figure*}
\includegraphics[width=0.49\linewidth]{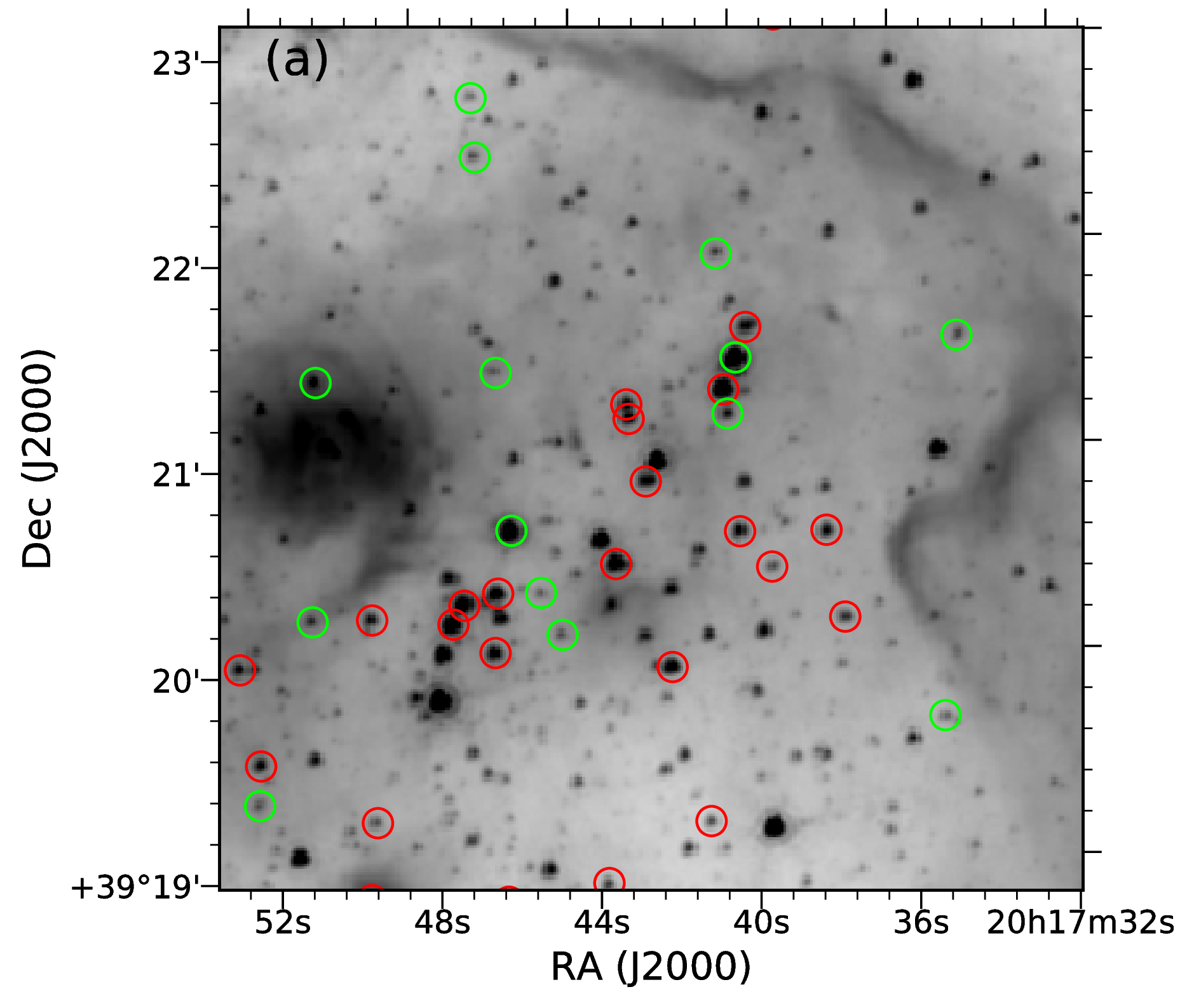}~\includegraphics[width=0.49\linewidth]{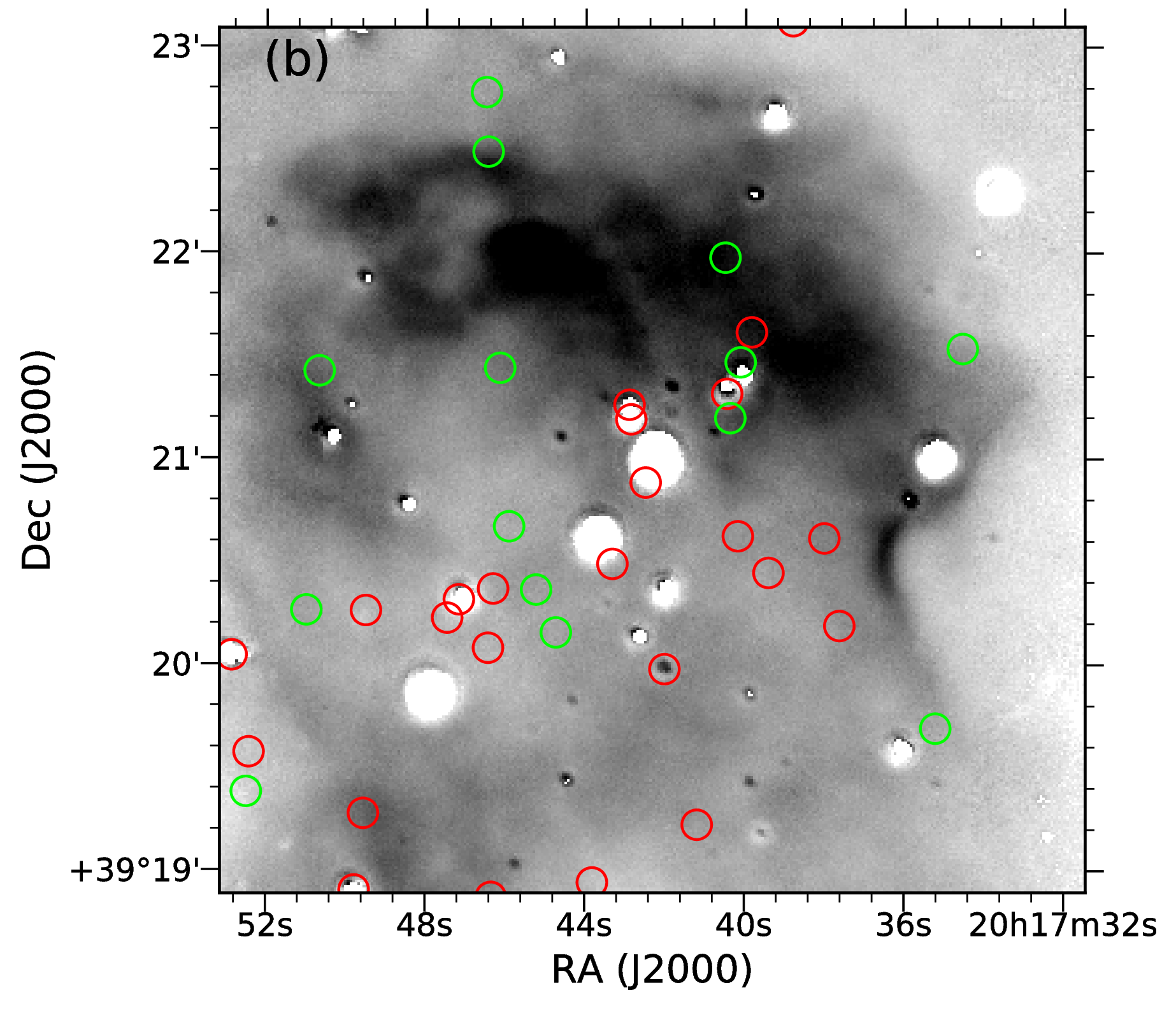}
\caption{Positions of selected YSOs inside the IR shell in the 3.6$\mu$m band
(a), and the \SII-emission distribution  (b) images. Class I protostars are
 shown in green, and Class II is in red.}\label{fig:YSOloc1}
\end{figure*}

\section{Young stellar objects towards the vdB~130 star cluster}

Four evolutionary stages can be distinguished among young stellar objects (YSOs). The objects corresponding to these stages are attributed
to Classes 0, I, II and III \citep{Lada1987,Lada2005}. Class 0 sources are very faint at wavelengths less than 10~$\mu$m, but have considerable
luminosities in the sub-millimetre range. Unlike these objects, those of Class I are much brighter in the IR range. However, it is impossible to
distinguish between these two classes of objects with only IR data available. Therefore, Class 0 protostars might be present among the selected
YSOs of Class I.

The objects of most interest to our investigation are the YSOs of Class I and
Class II. To separate them, we used the results of photometry in four IRAC
bands, carried out based on the \textit{Spitzer} CygnusX Legacy Project
\citep{cygX} survey. The catalog with selected point sources and their
photometry is available on the project
website\footnote{http://irsa.ipac.caltech.edu/cgi-bin/Gator/nph-dd}.

According to the point source catalog, 59 sources with measured magnitudes
in 3.6, 4.5 and 5.8 $\mu$m bands\footnote{Note that many of the remain
sources in catalogue have no listed magnitudes in 8 $\mu$m band. We
leave them for analysis because this band is not used in the identification
of the Class I objects. The number of identified Class II objects does not
change if we exclude these sources.} are observed in the studied
$21\arcmin\times17\arcmin$ region around the vdB 130 cluster. To single out the
Class I and II YSOs, we used the method proposed by \citet{gut08}. First, we
excluded the extragalactic sources (241 such sources were discovered) and
unresolved shock knots (3 source) based on their colours (see details in sec.
4.1 and appendix of \citealt{gut08}). We then selected 69 Class I objects
(shown in blue) and 176 Class II objects (red) based on their positions on
the colour-colour diagrams (Fig.~\ref{fig:YSOsel}). The remaining sources are
either field stars or Class III objects.


Fig. \ref{fig:YSOloc} shows the $350 \mu$m, $8 \mu$m and $24 \mu$m band
positions of Class I and II YSOs. Additionally, we plotted the isocontours of
the molecular cloud on the $24 \mu$m band image of the region \citep{sch07}.
Two bursts of star formation are clearly seen: inside the IR shell and near
the dense cloud core in the western part of the region. We can see that star
formation is ongoing in the vicinity of vdB 130, in the remnants of the
progenitor cloud inside the IR shell. Moreover, the protostars are
distributed more or less evenly inside the IR shell, with an insignificant
concentration towards the bright features in the `Herschel cloud', which
coincides with the head of the molecular cloud (Fig.~\ref{fig:YSOloc}a). The
process of star formation has recently began in the second star forming
region, and here we observe a compact cluster of protostars near the dense
cloud core \citep{motte}.

The distribution of YSOs inside the IR shell is shown in Fig.~\ref{fig:YSOloc1}. Note that only rare individual Class I
protostars are observed in the area of enhanced optical emission between Blob W and the north-western part of the IR shell
(Fig.\ref{fig:reg.A0} top). The detailed analysis of the distribution of protostars and young stars in the area will be presented elsewhere.

\begin{figure*}
\includegraphics[width=\linewidth]{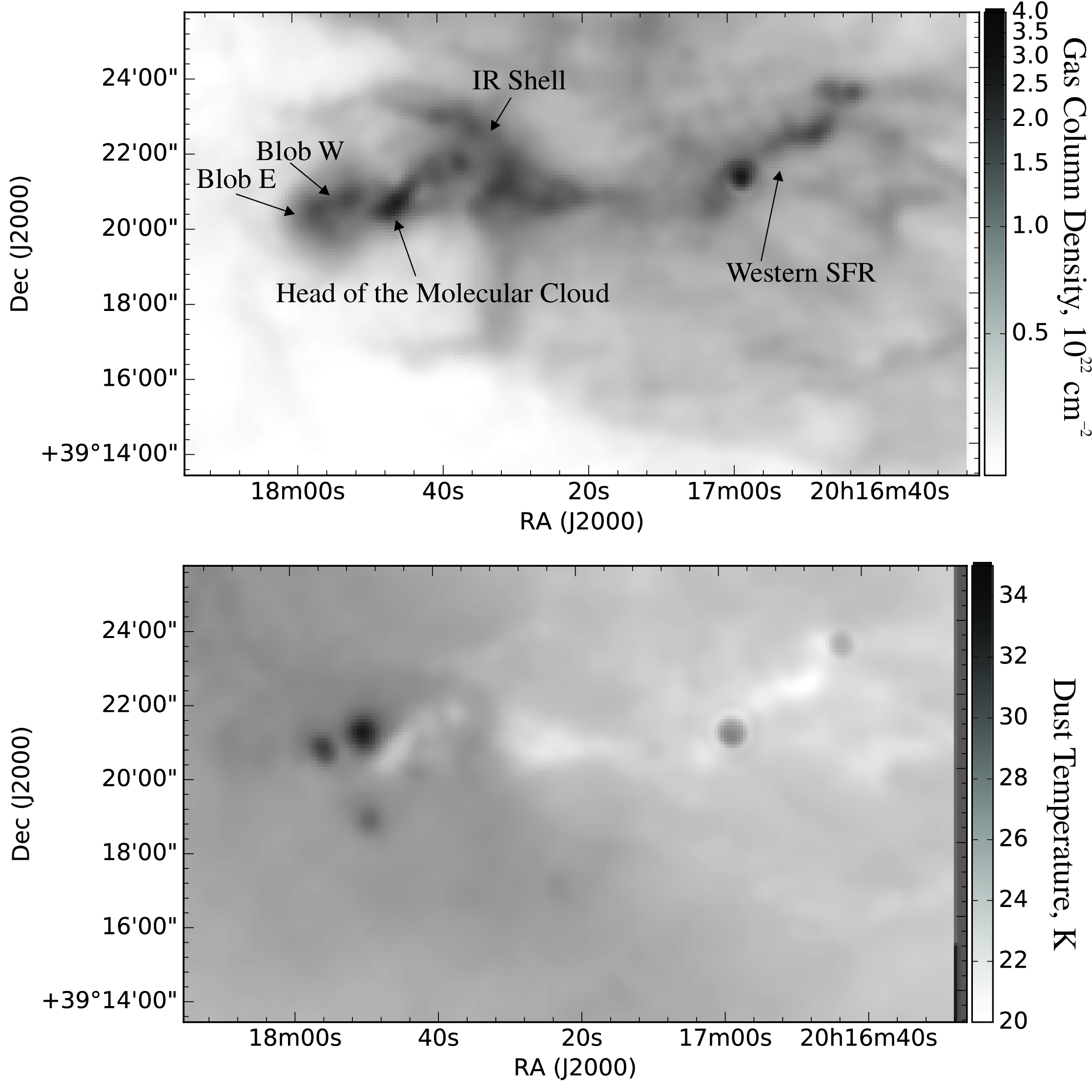}
\caption{Gas column density (upper panel) and dust temperature (bottom panel)
distributions in the large-scale vdB~130 vicinity.}\label{TdNd}
\end{figure*}

Finally, we have used {\em Herschel} data to analyse dust temperatures and
column densities in the studied large-scale region. Specifically,
observational data in 70, 160, 250, and 350~$\mu$m {\em Herschel} bands were
fitted to the modified black body to estimate $T_{\rm d}$ and $N_{\rm d}$, as
shown in Fig.~\ref{TdNd}. In the upper panel we clearly see the dense
vicinity of Blob W and Blob E, the head of the molecular cloud, and the IR
shell. The head of the molecular cloud appears to a part of a large filament
extending deep into the wall of the supershell. The second star-forming
region is located in one of the density enhancements along the filament.
While warm dust surrounding the blobs is clearly visible in the bottom panel
of Fig.~\ref{TdNd}, material is much colder in the areas of younger
star-forming regions. Apparently, protostar emission there still was not able
to heat large amounts of dust.

While the eastern star-forming region (in the head of the molecular pillar) might have been brought to life by the energetic influence of vdB 130 stars, the western star-forming region may represent an example of spontaneous star formation in the filament. In general, vdB~130 cluster, the dust blobs, and the eastern star-forming region may all be fossils of the filament, which is being gradually destroyed by the expansion of the Cyg OB1 supershell.

\section{Summary}

We investigated a region of ongoing star formation in the vicinity of the young star cluster vdB~130 located inside a
wall of an expanding supershell around the Cyg OB1 association. A complex of gas and dust nebulae around the stars of the vdB~130 cluster
is observed in a 3~pc region. From north, west, and south the complex is surrounded by an IR shell, visible in the 3.6 -- 24~$\mu$m near IR range, and is open from the side of the association. The IR shell represents an ionization front generated by the vdB~130 stars, which propagates through the cometary molecular cloud (pillar), formed by the wind and UV radiation of Cyg~OB1 stars. A shock front is observed in the eastern part of the region, where the IR shell is not seen. It is marked by the $I($\SII\,$6717,6731)/I($\Ha$)> 0.4$ ratio and coincides with the \SII\ filament in the eastern part of the area \citepalias{sit15}.

The following results were obtained.

\begin{enumerate}

\item  We analysed the \SII\ line radial velocity field inside the IR shell.
We show that the gas line-of-sight velocity in this area varies from $-30
\kms$ to $6 \kms$. The radial velocity, corresponding to large-scale motions
(galactic rotation) amounts to $-1 \div +6 \kms$ and coincides with the
velocity of the cometary molecular cloud. We confirm the suggestion mentioned
in \citetalias{sit15} about the expansion of the \HII\ region around the
cluster core stars with a velocity of about $15~\kms$, which is consistent
with the presence of the faint shock visible between Blob W and Blob E.

\item An analysis of the kinematics in the cluster region revealed a compact region Vel-A inside the IR shell,
where high-velocity motions are observed. Region Vel-A is projected onto a
$0.35\times0.15$~pc cavity seen in the optical emission distribution. The
ionized gas inside the cavity moves at an average velocity of $30 \kms$.
{Since no Class I or II protostars or stars
	brighter than 23~mag are found in this region, we suggest two explanations for the supersonic motions:
	the stellar wind from the point source, which is seen inside the cavity in \OIII\ line
	and near-IR images of the region, and a bipolar outflow from the protostellar disc of a Class 0 protostar, since
	region Vel-A appears to consist of two lobes}.

\item Another region (Vel-B) with relatively large negative gas radial velocities (in the range of $-~10 \div +1 \kms$) is
located in the IR shell. The observed increased velocities here may be a
consequence of the gas flow around the protrusion in the IR-shell.

\item  We discovered two star forming regions: a region of ongoing star formation in the vicinity of the vdB~130 cluster
inside the IR shell (around the pillar), and a star burst inside the cometary
molecular cloud, where a compact proto-cluster is observed near a dense
clump.

\item Temperatures and column densities in the studied region indicate that the head of the molecular cloud is a part of a large filament extending deep into the wall of the supershell. The cluster vdB~130, the dust blobs, and the entire eastern star-forming region may all be fossils of the filament, which is being gradually destroyed by the expansion of the Cyg OB1 supershell.

\end{enumerate}

\section*{Acknowledgements}

We thank the anonymous referee for useful comments that helped us to improve this manuscript.

This work was supported in part by M.V. Lomonosov Moscow State University
Program of Development. This work was supported by the Russian Foundation for
Basic Research (project \# 18-02-00976). The work is based in part on
observations obtained with the 6-m telescope of the Special Astrophysical
Observatory of the Russian Academy of Sciences. The authors thank S.A. Lamzin
for useful discussions and G.V. Smirnov-Pinchukov for his help with the data
reduction. {OE, TL and AT acknowledge the support from the Program of development of M.V. Lomonosov Moscow State University (Leading Scientific School `Physics of stars, relativistic objects and galaxies'). DW was supported by the P-12 Program (Presidium of the Russian Academy of Sciences) `Questions of
origin and evolution of the Universe'. }

This work is based in part on observations made with the {\it Spitzer} Space
Telescope, which is operated by the Jet Propulsion Laboratory, California
Institute of Technology under a contract with NASA and on observations made
with the {\it Herschel}, which is an ESA space observatory with science
instruments provided by European-led Principal Investigator consortia and
with important participation from NASA.

{This research has made use of `Aladin sky atlas' developed at CDS, Strasbourg Observatory, France \citep{2000A&AS..143...33B,2014ASPC..485..277B}.}

\appendix

\section{Region map} \label{rmap}

To facilitate reading the paper, we provide a sketch of the main features in the cluster vdB 130 area.

\begin{figure*}
\includegraphics[width=\linewidth]{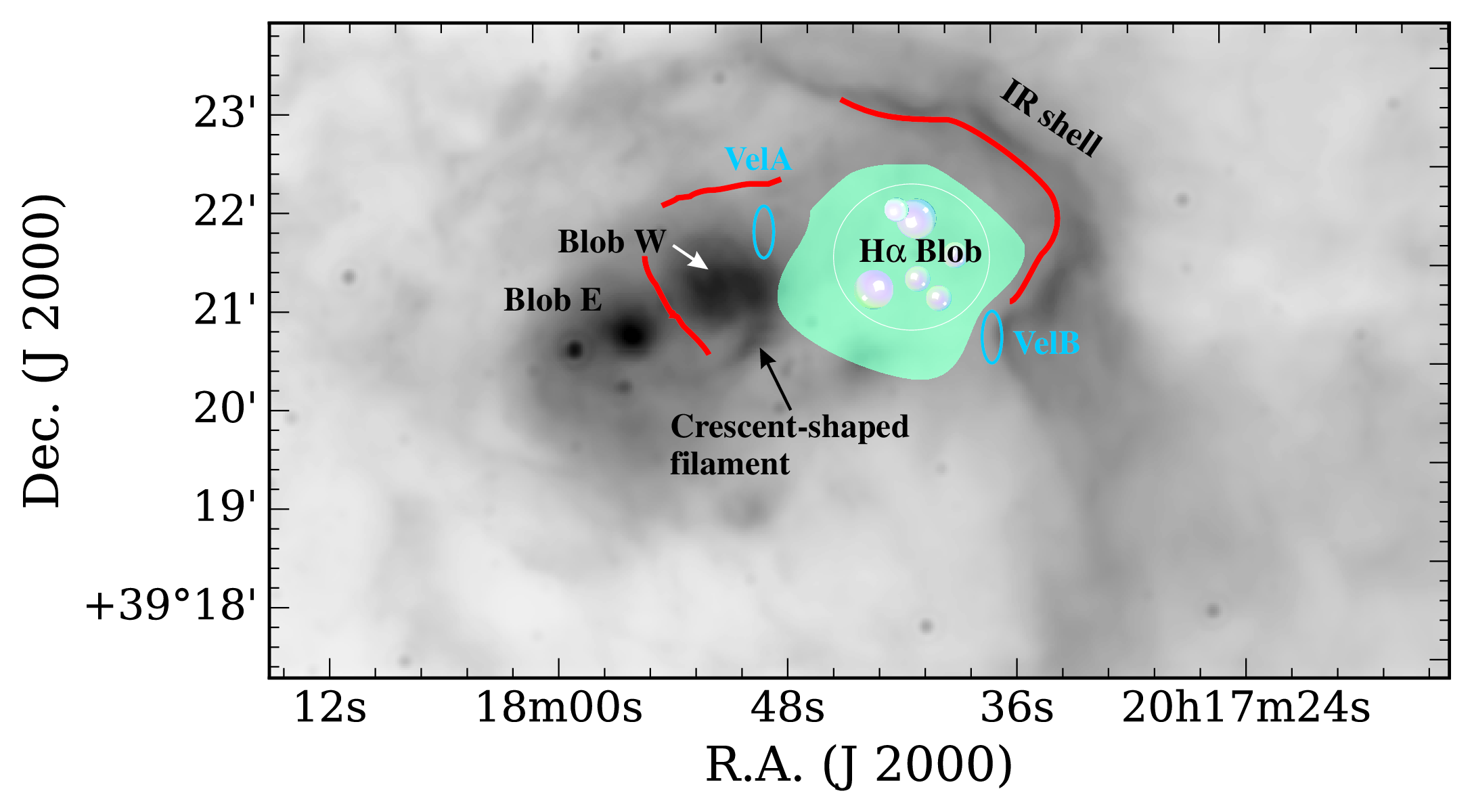}

\caption{A sketch of the main features in the vdb 130 area. The white
circle indicate a position of the \Ha\ Blob around the cluster core stars. A blue shaded area shows the region of the bright \OIII\ emission. Blue contours outline the regions with high velocity motions. Red curves outline approximate locations of the ionization front on the west and shock front to the east of the IR shell. Note that the IR shell has been
designated as Big Filament in \citetalias{sit15}.}\label{mapa}
\end{figure*}

\bsp    
\label{lastpage}
\end{document}